\documentclass[twocolumn,epjc3]{svjour3}  
\smartqed

\usepackage{graphicx}
\usepackage{amsmath,amssymb}
\usepackage{booktabs}
\usepackage{microtype}
\usepackage[colorlinks=true,allcolors=blue]{hyperref}

\journalname{Eur. Phys. J. C}

\newcommand{\ve}{\varepsilon}
\newcommand{\etr}{\varepsilon_{\rm tr}}
\newcommand{\ptr}{p_{\rm tr}}
\newcommand{\ntr}{n_{\rm tr}}
\newcommand{\Msun}{M_\odot}
\newcommand{\MTOV}{M_{\rm TOV}}
\newcommand{\dmax}{\Delta\varepsilon_{\max}}

\begin{document}
\sloppy

\title{Causality and perturbative-QCD constraints on first-order phase transitions in neutron stars}

\author{Asim Kumar Saha\thanksref{e1,addr1}
        \and
        Ritam Mallick\thanksref{e2,addr1}
}
\thankstext{e1}{e-mail: asim21@iiserb.ac.in}
\thankstext{e2}{e-mail: mallick@iiserb.ac.in}

\institute{Department of Physics, Indian Institute of Science Education and Research Bhopal, Bhopal, Madhya Pradesh 462066, India \label{addr1}
}

\date{Received: date / Accepted: date}

\maketitle

\begin{abstract}

Causality bounds the compactness and maximum mass of neutron stars. We generalise this bound to equations of state (EOSs) with a first-order phase transition between maximally stiff phases. In units of the transition energy density, the problem depends only on $q=\ptr/\etr$ and $j=\Delta\ve/\etr$. The maximum compactness equals the smooth value $\mathcal{C}=0.354$ for $q\to0$ or $j\to0$, is unaffected for $q\gtrsim0.67$, and is otherwise reduced by typically $\lesssim0.04$ when the maximum-mass star is a hybrid star and by up to $0.13$ when it is the last hadronic star. Matching to a Baym--Pethick--Sutherland crust and a chiral-EFT polytrope at $n_m=1.1\,n_0$ and inverting the bound, no causal EOS supports $2.01\,\Msun$  if the energy-density jump exceeds $\simeq460$--$495~{\rm MeV\,fm^{-3}}$. Additionally, imposing radius constraint of $R\ge11.6$~km it shrinks to $\simeq235$--$395~{\rm MeV\,fm^{-3}}$. These limits hold whatever the matter above the transition, for onsets within $\simeq0.2\,n_0$ of $n_m$. Relaxing maximal stiffness, we build EOSs with a transition at $\mu_{\rm tr}=1000$--$1300$~MeV, matched to perturbative QCD (pQCD) at $2.6$~GeV. pQCD limits the jump; EOSs compatible with PSR J0740+6620, J1614$-$2230, J0437$-$4715, J0614$-$3329, GW170817, and XTE J1814-338 appear only at $\mu_{\rm tr}=1200$ and $1250$~MeV; and inside the causal window, the largest jump agrees with the bound within the grid resolution. Finally, inserting a transition at $2.01\,\Msun$ into hadronic EOSs that satisfy five pulsar observations, the EOS with the lowest break density gives a narrow stable hybrid window ($\Delta M\simeq0.005\,\Msun$), compatible with XTE J1814-338 for jumps of $6$--$8\,n_0$ and with pQCD; three EOSs with higher break densities give none.

\keywords{Causality \and Equation of state \and First-order phase transition \and Neutron stars \and  Twin stars}
\end{abstract}

%=============================================================================
\section{Introduction}
\label{sec:intro}
%=============================================================================

The interiors of neutron stars (NSs) probe cold matter at a few times the nuclear saturation density $n_0=0.16~{\rm fm^{-3}}$, a regime inaccessible to terrestrial experiments and to first-principles calculations in quantum chromodynamics (QCD)~\cite{Fukushima2011,Baym2018}. Chiral effective field theory (EFT) constrains the equation of state (EOS) up to about $1$--$2\,n_0$~\cite{Hebeler2010,Tews2013,Hebeler2013}, and perturbative QCD (pQCD) becomes reliable only at densities far above those in NS cores~\cite{Kurkela2010}. In between, the EOS is constrained mainly by NS observations: the masses of PSR J0348+0432, J0740+6620 and J0952$-$0607~\cite{Antoniadis2013,Saffer:2024tlb,Fonseca2021,Romani2022}, radius measurements by the Neutron Star Interior Composition Explorer (NICER)~\cite{Riley2021,Miller2021,Salmi2024,Mauviard:2026gzc}, and the tidal deformability from GW170817~\cite{Abbott2018}.

The most robust theoretical input in this regime is causality, $c_s^2=\partial p/\partial\ve\le1$. Rhoades and Ruffini~\cite{Rhoades1974} showed that matching a known low-density EOS to maximally stiff matter ($c_s^2=1$) yields the largest possible maximum mass. This argument was refined in Refs.~\cite{Hartle1978,Kalogera1996,Koranda1997} and summarised by Lattimer~\cite{Lattimer2012}. A one-parameter family of maximally compact EOSs, $p=\ve-\ve_0$, yields maximum-mass configurations that all have the same compactness, $\mathcal{C}\equiv GM/(Rc^2)=0.354$. Their masses scale as $M_{\max}\simeq4.1\,(\ve_s/\ve_0)^{1/2}\Msun$, where $\ve_s$ is the saturation energy density. The locus of these configurations in the mass--radius ($M$--$R$) plane is the ``causality line''. These bounds hold for \emph{any} EOS that coincides with the known one up to the matching density, and that is why they are useful.

If the transition from hadronic to quark matter at low temperature is a first-order phase transition (FOPT)~\cite{Stephanov2004,prasad2022,gorda2023_b,thakur2024,Verma2025,verma2025b}, the EOS contains a segment with $c_s^2=0$ and a discontinuity $\Delta\ve$ in energy density at constant pressure. Such transitions can produce disconnected hybrid-star branches and ``twin'' stars~\cite{Gerlach1968,Glendenning2000,Alford2013,Christian2018,sedrakian2020,sedrakian2023}. Whether the hybrid branch connects to the hadronic one is governed by the Seidov criterion~\cite{Seidov1971}. Further the robustness of Seidov criterion has been studied in \cite{Alford2013,montana2018,atharva2026}. Maximally stiff cores with a density jump have been considered before in the context of the maximum mass~\cite{Zdunik2013,Chamel2013a,Chamel2013b}, while the stability of hybrid branches in a constant-speed-of-sound parametrisation that includes $c_s^2=1$ has been shown in Ref.~\cite{Alford2013}.

In this work we address three points. First, we formulate the problem in terms of two dimensionless numbers, $q=\ptr/\etr$ and $j=\Delta\ve/\etr$. We show explicitly that both limits recover the smooth bound, and we map the reduction of the maximum compactness over the full $(q,j)$ plane. Second, because maximal stiffness above the matching density maximises the mass, inverting the condition that the maximum mass $\MTOV$ of the Tolman--Oppenheimer--Volkoff (TOV) solutions equals the observed mass, $\MTOV=M_{\rm obs}$, gives an upper limit on the transition strength, $\dmax(\ntr)$. This limit holds for \emph{any} causal EOS sharing the low-density part, irrespective of the matter above the transition. We quantify how far such limits reach, and how much they tighten when a lower radius limit is imposed. Third, we turn to the observation of XTE J1814-338~\cite{Kini:2024ggu} (J1814-338 below), a source with a small mass and radius, which makes it very compact and puts it in tension with the NICER observations of PSR J0030+0451, PSR J0437$-$4715, PSR J0614$-$3329, PSR J0740+6620 and PSR J1614$-$2230 (J0030, J0437, J0614, J0740 and J1614 below). One possible way out is a first-order phase transition: the hadronic branch satisfies the NICER observations, while the second, hybrid branch can be made to satisfy XTE J1814-338. We therefore comment on the transition parameters, within the admissible range, for which this is feasible.

The causal construction describes the outer envelope of the allowed EOSs. Realistic EOSs are restricted further at high density, where the pressure of cold quark matter is known from pQCD~\cite{Kurkela2010,Gorda2021}, and where thermodynamic consistency and causality impose constraints at NS densities~\cite{Komoltsev2022,Gorda2023}. Interpolating the speed of sound between the nuclear and the pQCD regime is a standard way to build EOSs that satisfy these requirements~\cite{Annala2020}, and EOSs with first-order transitions built in this way were compared in Ref.~\cite{Verma2025}. We therefore also construct such EOSs, with a first-order transition at a chemical potential $\mu_{\rm tr}$ and a density jump $\Delta n$, matched to pQCD at $\mu=2.6$~GeV. We use them to determine which transitions are compatible with pQCD and with the available observations, and to test the causal bound.

The paper is organised as follows. Section~\ref{sec:formalism} presents the formalism: the EOS, the scale-free variables, the junction conditions and the integration scheme, and the low-density EOS. The results are presented in Sect.~\ref{sec:results}. They consist of the scale-free results (Sect.~\ref{sec:scalefree}), the bounds obtained with a realistic low-density EOS (Sect.~\ref{sec:crust}), EOSs with a speed-of-sound interpolation that are matched to pQCD and compared with the observations and with the bound (Sect.~\ref{sec:sosi}), and stable twin branches that reach the compact source J1814-338 (Sect.~\ref{sec:twins}). The discussion and summary are presented in Sect.~\ref{sec:discussion_summary}. We use $G=c=1$ unless stated otherwise.

%=============================================================================
\section{Formalism}
\label{sec:formalism}
%=============================================================================

\subsection{Equation of state}

\begin{figure}
\centering
\includegraphics[width=\columnwidth]{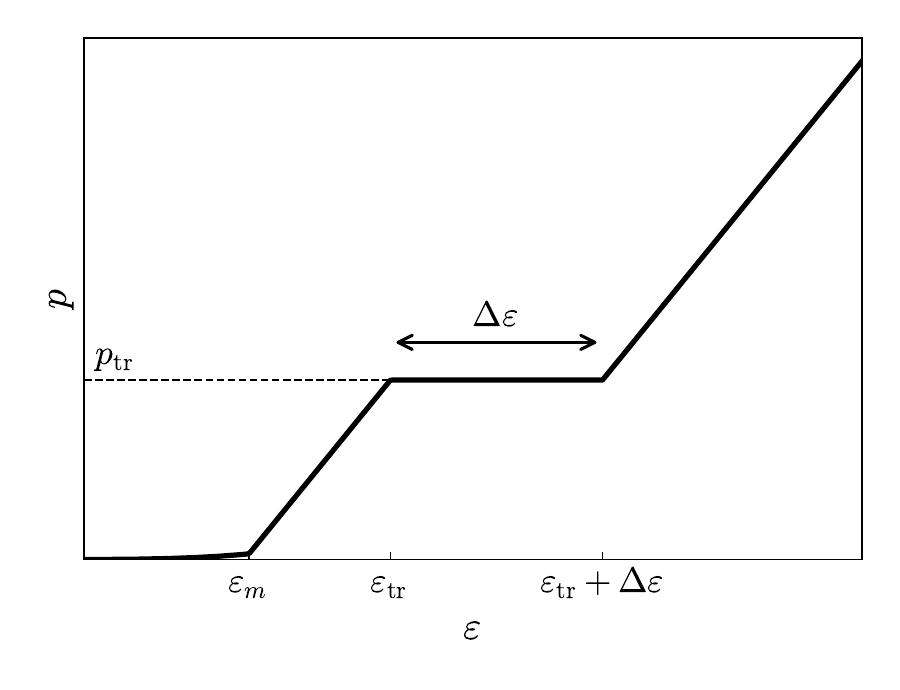}
\caption{Schematic EOS. A known low-density EOS is matched at $\ve_m$ to maximally stiff matter ($c_s^2=1$). A first-order transition at pressure $\ptr$ produces an energy-density jump $\Delta\ve$ from $\etr$ to $\etr+\Delta\ve$, after which the EOS is again maximally stiff. In the scale-free case the low-density part is absent and $p=0$ for $\ve\le\ve_m$.}
\label{fig:eos}
\end{figure}

Above a matching point $(\ve_m,p_m)$ we take the EOS to be maximally stiff up to the transition, to jump by $\Delta\ve$ at constant pressure $\ptr$, and to be maximally stiff again above it (Fig.~\ref{fig:eos}). Written as $\ve(p)$, this reads
\begin{equation}
\ve(p)=
\begin{cases}
\ve_{\rm low}(p), & p<p_m,\\
p+a, & p_m\le p\le\ptr,\\
p+a+\Delta\ve, & p>\ptr,
\end{cases}
\qquad a\equiv\ve_m-p_m .
\label{eq:eos}
\end{equation}
The EOS is continuous in pressure at $p_m$ by construction, and $\etr=\ptr+a$. The baryon density on a stiff segment follows from the first law, $d n/n=d\ve/(\ve+p)$, which gives
\begin{equation}
n(\ve)=n_m\left[\frac{2\ve-a}{\ve_m+p_m}\right]^{1/2},\qquad \ve_m<\ve\le\etr ,
\label{eq:nofe}
\end{equation}
and fixes the onset density $\ntr$ for a given $\etr$.

In the scale-free case there is no low-density part. Then $p_m=0$ and $a=\ve_m$, and the surface of the star has the finite energy density $\ve_m$, exactly as for the maximally compact EOS of Refs.~\cite{Koranda1997,Lattimer2012}.

\subsection{Scale-free variables}

Measuring all quantities in units of $\etr$, with
\begin{equation}
\tilde\ve=\frac{\ve}{\etr},\quad \tilde p=\frac{p}{\etr},\quad x=r\sqrt{\etr},\quad y=m\sqrt{\etr},
\end{equation}
the scale-free EOS becomes
\begin{equation}
\tilde\ve(\tilde p)=
\begin{cases}
\tilde p+1-q, & 0\le\tilde p\le q,\\
\tilde p+1-q+j, & \tilde p>q,
\end{cases}
\label{eq:eos_qj}
\end{equation}
with
\begin{equation}
q=\frac{\ptr}{\etr}\in[0,1),\qquad j=\frac{\Delta\ve}{\etr}\ge0 .
\end{equation}
The TOV equations~\cite{Tolman1939,Oppenheimer1939},
\begin{align}
\frac{d\tilde p}{dx}&=-\frac{(\tilde\ve+\tilde p)\,(y+4\pi x^3\tilde p)}{x\,(x-2y)},\label{eq:tov1}\\
\frac{dy}{dx}&=4\pi x^2\,\tilde\ve(\tilde p),\label{eq:tov2}
\end{align}
contain no other parameter, so every scale-free quantity is a function of $(q,j)$ and of the central pressure alone. Physical masses and radii follow from
\begin{align}
M&=48.1\,Y\left(\frac{150~{\rm MeV\,fm^{-3}}}{\etr}\right)^{1/2}\Msun,\nonumber\\
R&=71.0\,X\left(\frac{150~{\rm MeV\,fm^{-3}}}{\etr}\right)^{1/2}{\rm km},
\label{eq:units}
\end{align}
where $X$ and $Y$ denote the scaled radius and mass of a given configuration. For the maximally compact EOS with $\ve_0=150~{\rm MeV\,fm^{-3}}$, Eq.~(\ref{eq:units}) gives $M_{\max}=4.09\,\Msun$, consistent with Ref.~\cite{Lattimer2012}.

For each EOS, $\MTOV$ is the largest mass on any branch that is stable against radial perturbations. We assume rapid conversion at the interface, so stability requires $dM/d\varepsilon_c>0$~\cite{Pereira2018}. The TOV equations are integrated with a fixed-step fourth-order Runge--Kutta (RK4) scheme. The stability of the second (hybrid) branches, and the largest central density up to which a sequence remains stable, are determined with the stable-branch finder described in \ref{app:branch-finder}.

\subsection{Low-density EOS}

For the realistic case we use the Baym--Pethick--Sutherland (BPS) crust~\cite{Baym1971}, continued above the crust by a chiral-EFT (CET) band that we represent by a polytrope,
\begin{equation}
p(n)=K\left(\frac{n}{n_0}\right)^{\Gamma_{\rm CET}},
\end{equation}
with $K$ fixed by pressure continuity at the upper edge of the crust. The polytrope is used up to the matching density $n_m=1.1\,n_0$, which lies within the range up to which chiral EFT is commonly regarded as reliable~\cite{Hebeler2013,Tews2018,Drischler2021}. In Sects.~\ref{sec:scalefree} and~\ref{sec:crust} we take the stiffest polytrope, $\Gamma_{\rm CET}=3.23$. It gives $p=3.80~{\rm MeV\,fm^{-3}}$ at $n_0$, and $p_m=5.17~{\rm MeV\,fm^{-3}}$ and $\ve_m=168.7~{\rm MeV\,fm^{-3}}$ at $n_m$. Softer polytropes ($\Gamma_{\rm CET}=1.77$ and $2.23$) change the bound of Sect.~\ref{sec:crust} by only $1$--$2\%$, and the stiffest polytrope gives the weakest bound. The three polytropes are used in Sect.~\ref{sec:sosi}.

%=============================================================================
\section{Results}
\label{sec:results}

\subsection{Scale-free results}
\label{sec:scalefree}
%=============================================================================

% \subsubsection{Limits}

Two limits of Eq.~(\ref{eq:eos_qj}) reduce exactly to the smooth maximally compact EOS. For $j=0$ there is no transition, and the EOS is $p=\ve-\ve_0$ with $\ve_0=(1-q)\etr$. For $q=0$ the transition occurs at zero pressure, and the EOS is again $p=\ve-\ve_0$, now with $\ve_0=(1+j)\etr$. In both cases
\begin{equation}
Y_{\max}=\frac{0.08513}{\sqrt{\ve_0/\etr}},\qquad X_{\max}=\frac{0.2404}{\sqrt{\ve_0/\etr}},\qquad \mathcal{C}=0.354 .
\label{eq:limits}
\end{equation}
Our integration reproduces these values to a relative accuracy better than $10^{-4}$, which we use as a test of the numerics. Hence both edges of the $(q,j)$ plane lie exactly on the smooth causality line. Any reduction of the maximum compactness by a FOPT must vanish as $q\to0$ or $j\to0$.

\begin{figure}
\centering
\includegraphics[width=\columnwidth]{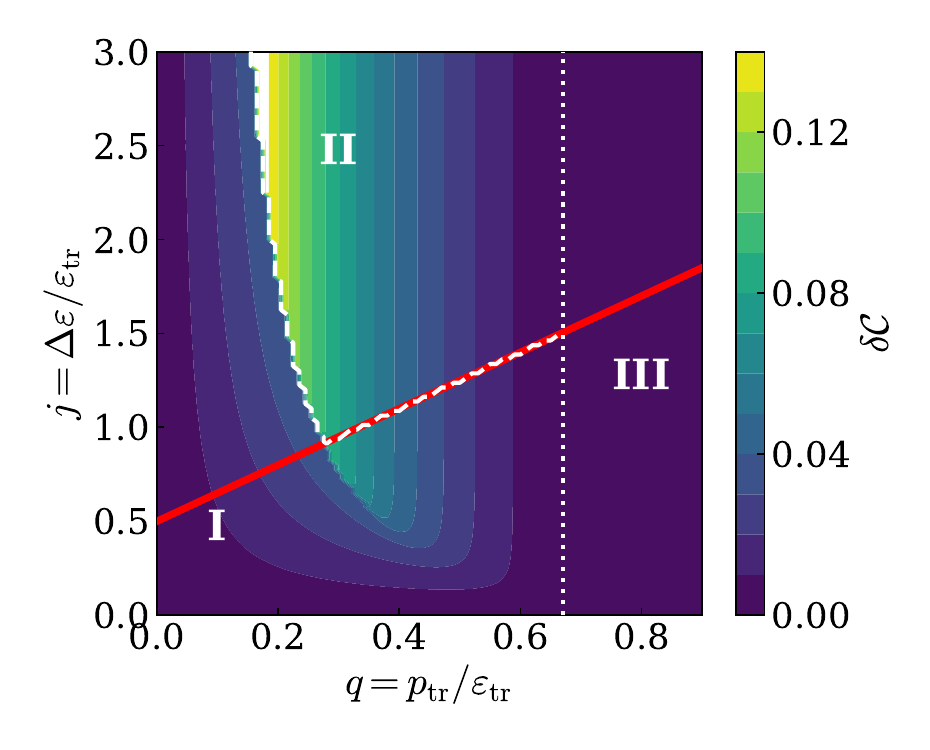}
\caption{Compactness deficit $\delta\mathcal{C}=0.354-\mathcal{C}_{\max}$ of the maximum-mass configuration as a function of the reduced transition pressure $q$ and strength $j$ (scale-free EOS). The red line is the Seidov criterion $j=\tfrac12+\tfrac32q$, above which the hybrid branch is disconnected from the hadronic one. The white dashed line separates region~I, where the maximum-mass star is a hybrid star, from region~II, where it is the last hadronic star and $\delta\mathcal{C}$ depends on $q$ only. In region~III ($q>0.670$) the transition lies above the central pressure of the maximum-mass star. Along $q=0$ and $j=0$ the smooth causal limit is recovered exactly.}
\label{fig:dbeta}
\end{figure}

We define the compactness deficit
\begin{equation}
\delta\mathcal{C}(q,j)\equiv0.354-\mathcal{C}_{\max}(q,j)\ge0 ,
\end{equation}
where $\mathcal{C}_{\max}$ is the compactness of the maximum-mass configuration. Figure~\ref{fig:dbeta} shows $\delta\mathcal{C}$ over $0\le q\le0.9$ and $0\le j\le3$. We find $\delta\mathcal{C}\ge0$ everywhere, to within the numerical accuracy ($|\delta\mathcal{C}|<5\times10^{-5}$ at the edges). The plane separates into three regions.

\paragraph{Region I.} The maximum-mass star is a hybrid star. The deficit grows with $j$ at fixed $q$, but remains moderate: typically $\delta\mathcal{C}\lesssim0.04$ ($\lesssim11\%$ of the smooth value), with values up to $\approx0.06$ close to the boundary of region~II. For example, $q=0.2$ and $j=1.5$ give $\delta\mathcal{C}=0.036$. This region includes disconnected hybrid branches (above the Seidov line, $j>\tfrac12+\tfrac32q$~\cite{Seidov1971}) as long as the third family is stable and heavier than the last hadronic star.

\paragraph{Region II.} The hybrid branch is either unstable or lighter than the end of the hadronic branch, and $\MTOV$ is the hadronic star with central pressure $p_c=\ptr$. Its compactness is that of a maximally compact star with $p_c/\ve_0=q/(1-q)$. It depends on $q$ alone, so $\delta\mathcal{C}$ is independent of $j$ and becomes large at small $q$: $\delta\mathcal{C}=0.13$ at $q=0.2$ and $0.025$ at $q=0.5$. The boundary between regions I and II lies close to $j\simeq1$ for $0.25\lesssim q\lesssim0.6$. It moves to larger $j$ as $q\to0$; for $q\le0.15$ the hybrid branch still carries $\MTOV$ at $j=3$, the largest value we considered. The limit $q\to0$ is therefore non-uniform in $j$. This is how a finite gap at $q>0$ coexists with the continuity of Eq.~(\ref{eq:limits}).

\paragraph{Region III.} For the maximally compact EOS the maximum-mass star has $p_c=2.03\,\ve_0$. With $\ve_0=(1-q)\etr$, the transition lies above the centre of the maximum-mass star whenever
\begin{equation}
q>q^*=\frac{2.03}{3.03}=0.670 .
\end{equation}
In this region the FOPT does not affect $\MTOV$ at all, and $\delta\mathcal{C}=0$ identically.

A FOPT with given $(q,j)$ reduces the maximum attainable compactness by $\delta\mathcal{C}(q,j)$. The reduction vanishes on the boundaries $q=0$, $j=0$ and in region III, so no gap exists independent of the transition parameters. Observationally, a star with compactness above $0.354-\delta\mathcal{C}(q,j)$ would exclude a transition with those parameters. With the presently measured compactnesses, $\mathcal{C}\lesssim0.27$, this does not yet exclude any part of the $(q,j)$ plane. A stronger use of causality is obtained by fixing the low-density EOS, as we do next.

%=============================================================================
\subsection{Realistic low-density EOS and the inverted bound}
\label{sec:crust}
%=============================================================================

% \subsubsection{Smooth reference}

Matching the BPS+CET EOS to $c_s^2=1$ matter at $n_m=1.1\,n_0$ without a transition gives the largest mass compatible with causality, $M_{\max}=3.92\,\Msun$ with $R=16.8$~km (Table~\ref{tab:smooth}). The compactness, $\mathcal{C}=0.345$, stays slightly below $0.354$ because of the soft low-density envelope.

\begin{table}
\caption{Maximum mass, radius and compactness for the BPS+CET EOS ($\Gamma_{\rm CET}=3.23$) matched to maximally stiff matter at $n_m=1.1\,n_0$, with no phase transition.}
\label{tab:smooth}
\centering
\small\setlength{\tabcolsep}{4pt}
\begin{tabular}{cccccc}
\toprule
$n_m/n_0$ & $p_m$ & $\ve_m$ & $M_{\max}$ & $R$ & $\mathcal{C}$ \\
 & (MeV\,fm$^{-3}$) & (MeV\,fm$^{-3}$) & ($\Msun$) & (km) & \\
\midrule
1.1 & 5.17 & 168.7 & 3.92 & 16.8 & 0.345 \\
\bottomrule
\end{tabular}
\end{table}

\begin{figure}
\centering
\includegraphics[width=\columnwidth]{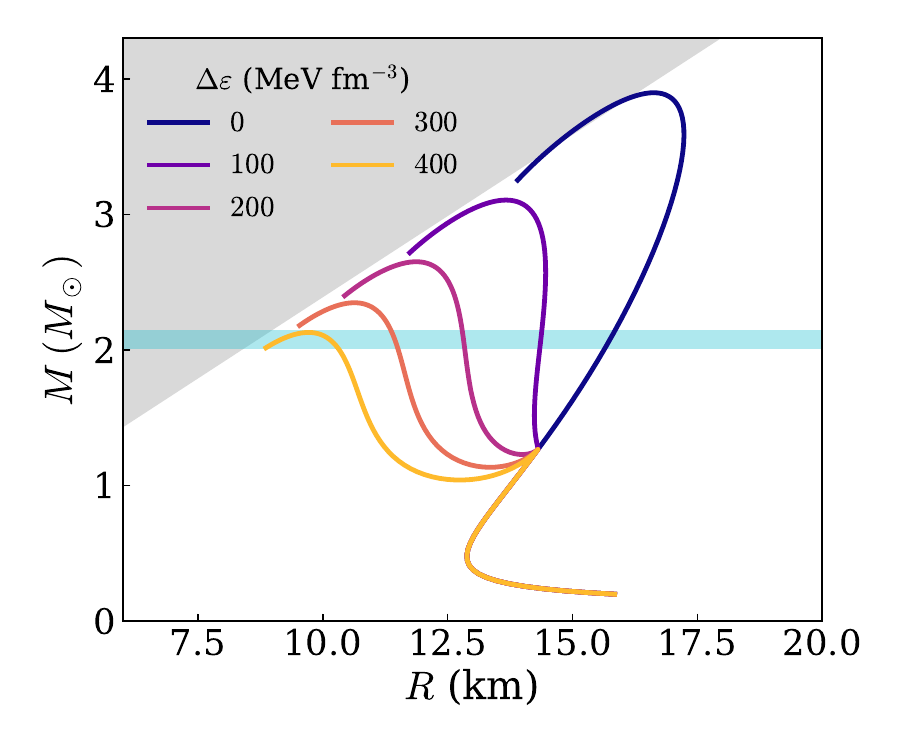}
\caption{Mass--radius sequences for the BPS+CET EOS ($\Gamma_{\rm CET}=3.23$) matched to maximally stiff matter at $n_m=1.1\,n_0$, with a FOPT at $\ntr=1.2\,n_0$ and the indicated $\Delta\ve$. The grey region is excluded by the smooth causality bound $\mathcal{C}\le0.354$. The horizontal band marks the mass of PSR J0740+6620~\cite{Fonseca2021}.}
\label{fig:mr}
\end{figure}

Figure~\ref{fig:mr} shows $M$--$R$ sequences for $n_m=1.1\,n_0$ and $\ntr=1.2\,n_0$. The hadronic branch ends at $1.45\,\Msun$. As $\Delta\ve$ increases, the hybrid branch is pushed to smaller radii and lower masses: $\MTOV=3.92$, $3.12$, $2.67$, $2.36$ and $2.14\,\Msun$ for $\Delta\ve=0$, $100$, $200$, $300$ and $400~{\rm MeV\,fm^{-3}}$. We find $\MTOV$ to decrease monotonically with $\Delta\ve$ at fixed $\ntr$ in all cases considered. This property makes the inversion below well defined.

% \subsubsection{Upper limit on the transition strength}

In this section we derive an upper limit on the strength of the transition from the measured mass of a massive pulsar. The low-density EOS is held fixed up to the matching density $n_m$. Above $n_m$ we consider the stiffest EOS that causality allows, with a first-order transition of a given set of parameters. This EOS defines a ceiling on the maximum mass that any causal EOS with the same low-density part and the same transition can reach. If the mass of an observed pulsar exceeds this ceiling, the transition with those parameters is excluded.

For a given onset density $\ntr$ and energy-density jump $\Delta\ve$, consider the EOS as described in Sect.~\ref{sec:formalism}. It follows the low-density EOS up to $n_m$, is maximally stiff ($c_s^2=1$) from $n_m$ to $\ntr$, jumps by $\Delta\ve$ at constant pressure, and is maximally stiff again above the jump. Its maximum mass is denoted $\MTOV(\ntr,\Delta\ve)$. Any other causal EOS that has the same low-density part and the same transition must have $c_s^2\le1$ everywhere above $n_m$, so it is softer than this one at some densities. A softer EOS cannot carry more mass, for the case without a transition \cite{Rhoades1974}. Hence $\MTOV(\ntr,\Delta\ve)$ is the largest maximum mass that \emph{any} causal EOS with these transition parameters can reach, regardless of the matter above the transition. A realistic EOS is therefore expected to have a maximum mass $M^{\rm real}_{\max}\le\MTOV(\ntr,\Delta\ve)$.

The ceiling becomes lower when the jump becomes larger. In Fig.~\ref{fig:mr} the onset is $\ntr=1.2\,n_0$, and the ceiling is $3.92\,\Msun$ without a jump and $2.67$ and $2.14\,\Msun$ for $\Delta\ve=200$ and $400~{\rm MeV\,fm^{-3}}$. We found this decrease with $\Delta\ve$ in all cases that we computed. A steady decrease guarantees that Eq.~(\ref{eq:inversion}) has a single solution, so that the allowed jumps form a single interval, $\Delta\ve\le\dmax$.

The observation of a pulsar of mass $M_{\rm obs}$ requires that the true EOS of dense matter supports such a star, $M^{\rm real}_{\max}\ge M_{\rm obs}$. Since $M^{\rm real}_{\max}\le\MTOV(\ntr,\Delta\ve)$, this implies
\begin{equation}
\MTOV(\ntr,\Delta\ve)\ge M_{\rm obs}.
\end{equation}
This is a necessary condition. It does not assume that the pulsar has the mass $\MTOV$; it only requires that the ceiling does not lie below the observed mass. If it does, not even the stiffest causal EOS can support the pulsar, and the transition is excluded. For example, at $\ntr=1.2\,n_0$ and $M_{\rm obs}=2.01\,\Msun$, a jump of $\Delta\ve=200~{\rm MeV\,fm^{-3}}$ is allowed, whereas a jump of $500~{\rm MeV\,fm^{-3}}$ is excluded, since the ceiling then falls below $2.01\,\Msun$.

Because the ceiling decreases with $\Delta\ve$, the condition is equivalent to an upper limit on the jump. This limit, $\dmax$, is the value at which the ceiling equals the observed mass,
\begin{equation}
\MTOV(\ntr,\dmax)=M_{\rm obs}.
\label{eq:inversion}
\end{equation}
All transitions with $\Delta\ve\le\dmax$ are compatible with the observation, and all transitions with $\Delta\ve>\dmax$ are excluded. The limit depends on the onset density, $\dmax=\dmax(\ntr)$, and is independent of the behaviour of matter above the transition.

The maximum mass $\MTOV$ of the stiff EOS is carried either by a purely hadronic star or by a star with a quark core. Denote by $M_1(\ntr)$ the mass of the last hadronic star, that is, the star whose central pressure has just reached the transition pressure, $p_c=\ptr$. All stars below this one are purely hadronic and are not affected by the jump. The hybrid stars, which have the transition inside, are the ones whose mass depends on $\Delta\ve$.

If $M_1(\ntr)\ge M_{\rm obs}$, the observed pulsar can already be a purely hadronic star. The condition above is then satisfied for any value of $\Delta\ve$, even for a very large jump, because the jump only affects stars above the last hadronic one. In this case a mass measurement by itself excludes nothing. If $M_1(\ntr)<M_{\rm obs}$, the observed star has to be a hybrid star, and then its mass is limited by the size of the jump, which gives the limit of Eq.~(\ref{eq:inversion}). The last hadronic star becomes heavier when the onset is moved to higher density, since the stiff hadronic segment between $n_m$ and $\ntr$ gets longer. There is therefore a threshold onset density $\ntr^*$, defined by
\begin{equation}
M_1(\ntr^*)=M_{\rm obs}.
\label{eq:nstar}
\end{equation}
For onsets $\ntr<\ntr^*$ the observed star must be a hybrid star and the limit $\dmax(\ntr)$ exists. For $\ntr\ge\ntr^*$ the stiffest possible hadronic matter reaches $M_{\rm obs}$ by itself and the mass places no bound on the transition. We note that this last statement is about the stiffest hadronic matter. A real, softer hadronic branch might not reach $M_{\rm obs}$, so for $\ntr>\ntr^*$ the mass measurement is silent and does not show that the transition is allowed.

\begin{figure}
\centering
\includegraphics[width=\columnwidth]{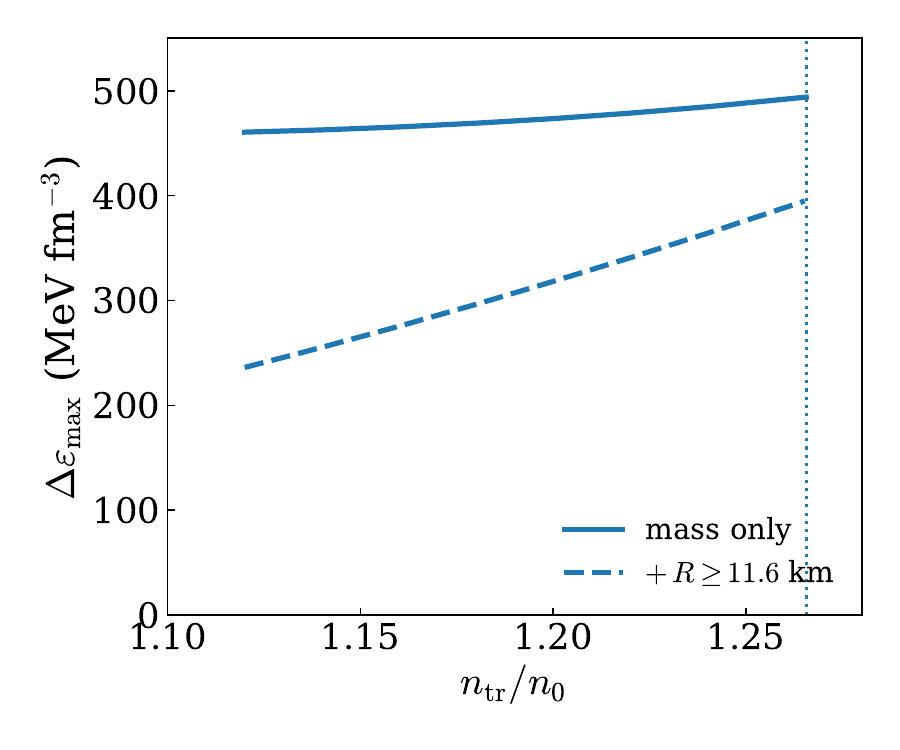}
\caption{Largest energy-density jump $\dmax$ compatible with a $2.01\,\Msun$ star, as a function of the transition onset $\ntr$, for the matching density $n_m=1.1\,n_0$. The solid curve uses the mass alone, Eq.~(\ref{eq:inversion}). The dashed curve additionally requires $R\ge11.6$~km for that star. Above the dotted vertical line ($\ntr^*$) the maximally stiff hadronic branch alone reaches $2.01\,\Msun$, and the mass places no bound on $\Delta\ve$.}
\label{fig:bound}
\end{figure}

Table~\ref{tab:bounds} and Fig.~\ref{fig:bound} give the upper limits on the strength of the transition for $M_{\rm obs}=2.01\,\Msun$, the lower edge of the $68\%$ credible interval of the mass of PSR J0740+6620, $M=2.08\pm0.07\,\Msun$~\cite{Fonseca2021}. For onsets between $\ntr=1.12\,n_0$ and the threshold $\ntr^*=1.27\,n_0$, the mass alone gives $\dmax=460$--$494~{\rm MeV\,fm^{-3}}$. In units of the energy density at the onset, this is a jump of about $2.5$--$2.7$ times $\etr$. If the radius of this star is required in addition to be at least $11.6$~km, the limit becomes stronger, $236$--$395~{\rm MeV\,fm^{-3}}$ (dashed curve in Fig.~\ref{fig:bound} and second row of Table~\ref{tab:bounds}). The value $11.6$~km is the lower edge of the $68\%$ credible interval of the NICER radius of PSR J0740+6620, $R=12.49^{+1.28}_{-0.88}$~km~\cite{Salmi2024}. We apply it to the $2.01\,\Msun$ star, which is slightly lighter than the star analysed there, and the derivation of the radius limit is given below. Heavier stars give stronger limits. For the central mass of PSR J0740+6620, $2.08\,\Msun$, the mass-only limit is $419$--$455~{\rm MeV\,fm^{-3}}$, and for $2.35\,\Msun$, the mass of PSR J0952$-$0607~\cite{Romani2022}, it is $293$--$337~{\rm MeV\,fm^{-3}}$. The behaviour of $\dmax(\ntr)$ for the three masses is shown in \ref{app:masses} (Fig.~\ref{fig:masses}). The threshold $\ntr^*$ lies only $0.17$--$0.22\,n_0$ above $n_m$ (Table~\ref{tab:bounds}), since maximally stiff nucleonic matter reaches $2\,\Msun$ already at very low central densities. The window of onset densities in which a mass measurement constrains a first-order transition through causality is therefore narrow. Both the window and the limits depend on the matching density $n_m$ and on the low-density EOS used up to it.

\begin{table}
\caption{Causal upper limits $\dmax$ on the energy-density jump (MeV\,fm$^{-3}$) for $n_m=1.1\,n_0$. The limit grows with the onset density $\ntr$, so it is given at the two ends of the range: ``low'' is the limit at the earliest onset computed, $\ntr=1.12\,n_0$, and ``high'' the limit as $\ntr$ approaches the threshold $\ntr^*$ from below. Here $\ntr^*$ is the onset above which the stiffest hadronic branch alone reaches $M_{\rm obs}$ and the mass gives no limit. The first row uses the observed mass $2.01\,\Msun$ alone, and the second row adds the requirement that the radius of this star be at least $11.6$~km. The last two rows give the mass-only limits for larger masses.}
\label{tab:bounds}
\centering
\begin{tabular}{cccc}
\toprule
& low & high & $\ntr^*/n_0$\\
\midrule
$M_{\rm obs}=2.01\,\Msun$ & 460 & 494 & 1.27\\
$2.01\,\Msun$ \& $R\ge11.6$\,km & 236 & 395 & 1.27\\
$M_{\rm obs}=2.08\,\Msun$ & 419 & 455 & 1.28\\
$M_{\rm obs}=2.35\,\Msun$ & 293 & 337 & 1.32\\
\bottomrule
\end{tabular}
\end{table}

% \subsubsection{Adding a lower radius limit}

Within the window $\ntr<\ntr^*$ the observed star must be a hybrid star. A lower limit $R_{\rm lo}$ on its radius then constrains $\Delta\ve$ further, provided that maximal stiffness also yields the largest radius at fixed mass. We tested this property numerically. Reducing $c_s^2$ to $0.7$ in the hadronic segment, the quark segment, or both, lowered the radius of the $2.01\,\Msun$ hybrid star in every case tested ($\ntr=1.15$, $1.2$ and $1.25\,n_0$, $\Delta\ve=100$ and $200~{\rm MeV\,fm^{-3}}$, at fixed transition pressure), by $0.02$--$0.61$~km (softening the quark segment leaves a purely hadronic star unchanged, as it must). We did not find a counterexample. With this property, $R(M_{\rm obs})\ge R_{\rm lo}$ implies $\Delta\ve\le\dmax^{(R)}$, where $\dmax^{(R)}$ solves $R_{\rm stiff}(M_{\rm obs};\ntr,\dmax^{(R)})=R_{\rm lo}$.

As an illustration we take $R_{\rm lo}=11.6$~km, the lower edge of the $68\%$ credible interval of the NICER radius of PSR J0740+6620, $R=12.49^{+1.28}_{-0.88}$~km~\cite{Salmi2024}. The resulting limit is shown as the dashed curve in Fig.~\ref{fig:bound}. Near the matching density the radius requirement reduces $\dmax$ by a factor of up to $1.9$, from $460$ to $236~{\rm MeV\,fm^{-3}}$ at $\ntr=1.12\,n_0$. The two limits approach each other as $\ntr\to\ntr^*$, where they are $494$ and $395~{\rm MeV\,fm^{-3}}$.

%=============================================================================
\subsection{Realistic EOSs: first-order transition, speed-of-sound interpolation and pQCD}
\label{sec:sosi}
%=============================================================================

The EOSs of Sects.~\ref{sec:scalefree} and~\ref{sec:crust} are extremal: they are maximally stiff everywhere above $n_m$ and decide what is \emph{allowed}. We now ask what remains when the matter above the transition has to connect to perturbative QCD (pQCD) at asymptotic densities. We construct such EOSs from the same ingredients, a low-density EOS, causal matter and a first-order jump, and add a speed-of-sound interpolation (SoSI) branch that is matched to the pQCD pressure at $\mu_{\rm end}=2.6$~GeV. Throughout this section $\mu=(\ve+p)/n$ is the baryon chemical potential, and the transition is specified by its chemical potential $\mu_{\rm tr}$ and the jump $\Delta n$ in baryon density, rather than by $\ve_{\rm tr}$ and $\Delta\ve$. The construction of these EOSs is described in \ref{app:construction}. In short, an EOS consists of the BPS crust and the CET polytrope up to $n_m$, a causal hadronic branch up to $\mu_{\rm tr}$, a first-order jump $\Delta n$ at constant $p$ and $\mu$, so that $\Delta\ve=\mu_{\rm tr}\Delta n$, a causal continuation up to the maximum-mass star, and a speed-of-sound interpolation matched to pQCD. We scan $41$ values of the jump, $\Delta n=0$--$3.2~{\rm fm^{-3}}$ in steps of $0.5\,n_0$, for $\mu_{\rm tr}=1000$, $1100$, $1200$, $1250$ and $1300$~MeV and for the three polytropes $\Gamma_{\rm CET}=1.77$, $2.23$ and $3.23$.

\begin{figure*}
\centering
\includegraphics[width=0.48\linewidth]{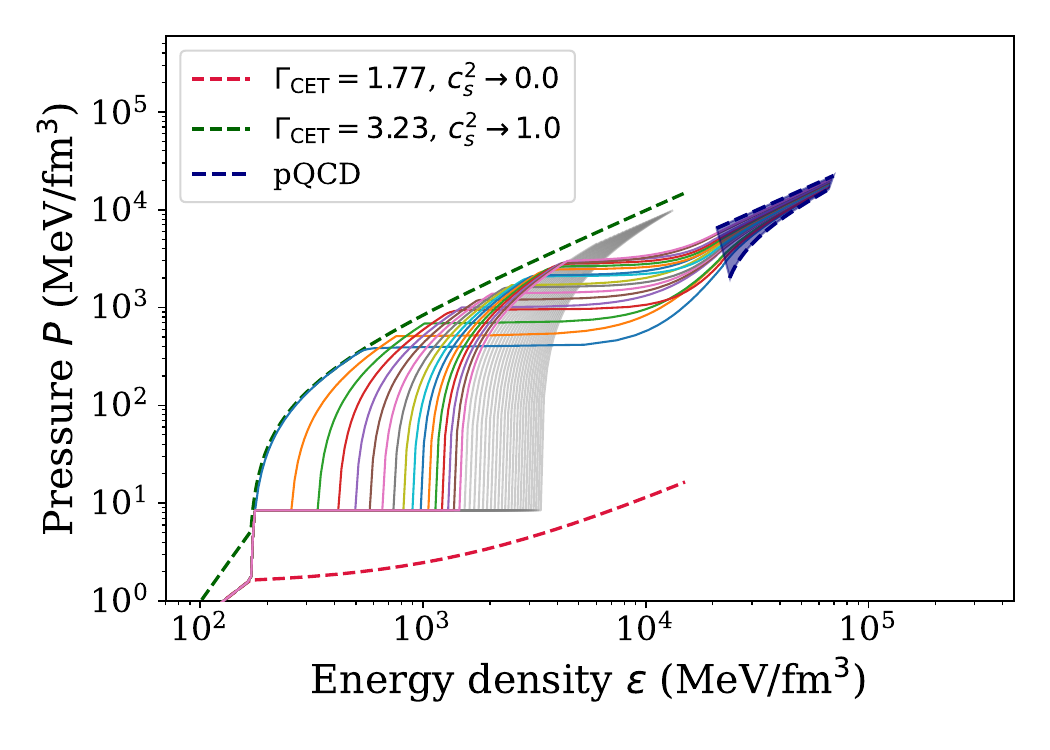}
\includegraphics[width=0.48\linewidth]{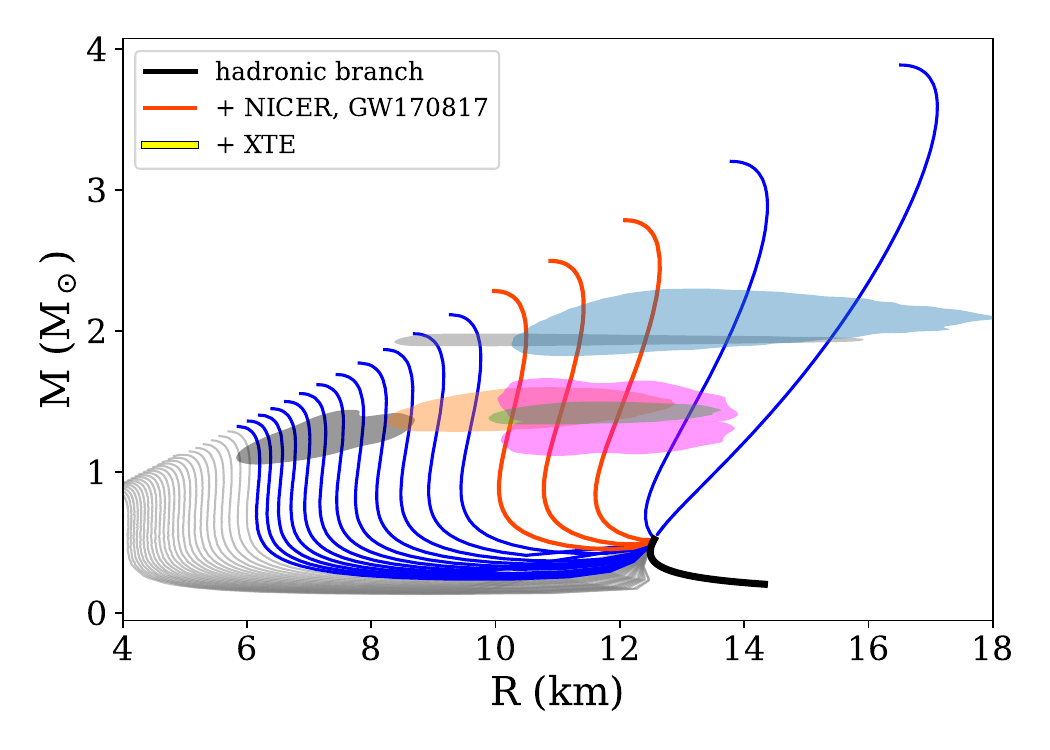}

\includegraphics[width=0.48\linewidth]{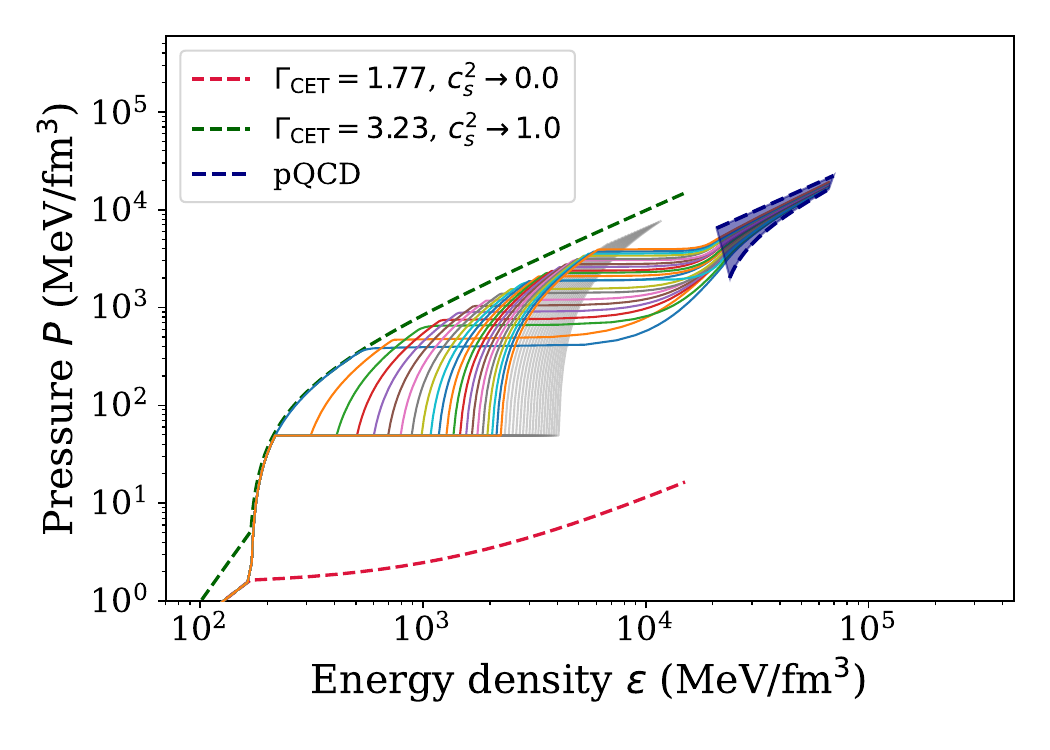}
\includegraphics[width=0.48\linewidth]{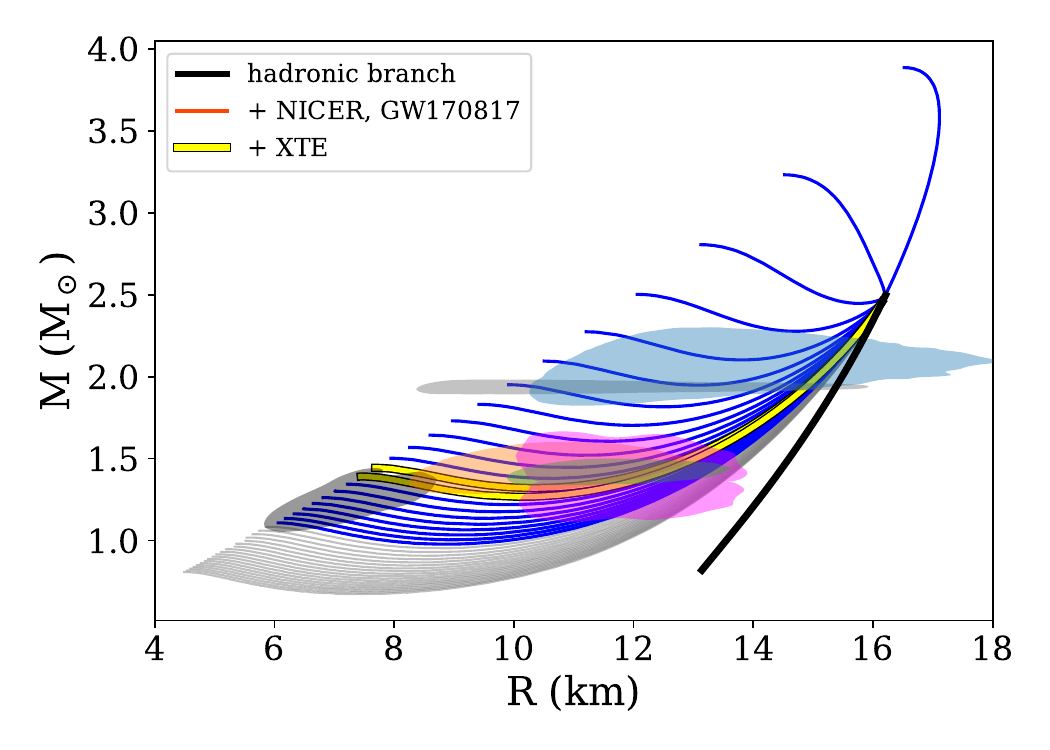}

\includegraphics[width=0.48\linewidth]{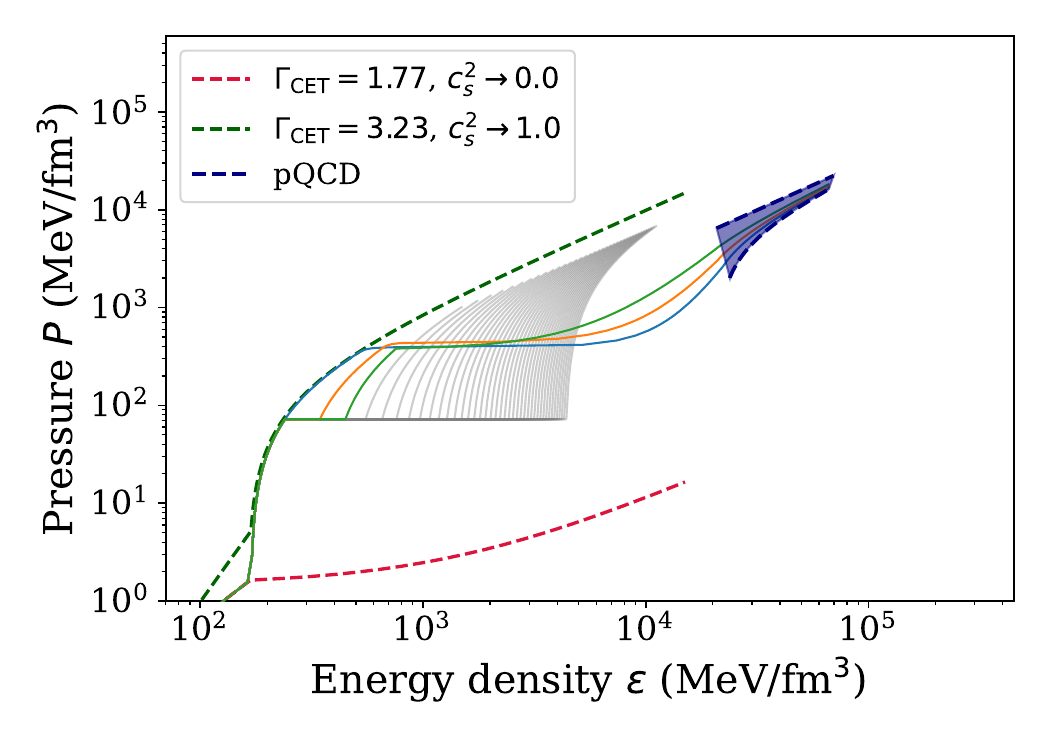}
\includegraphics[width=0.48\linewidth]{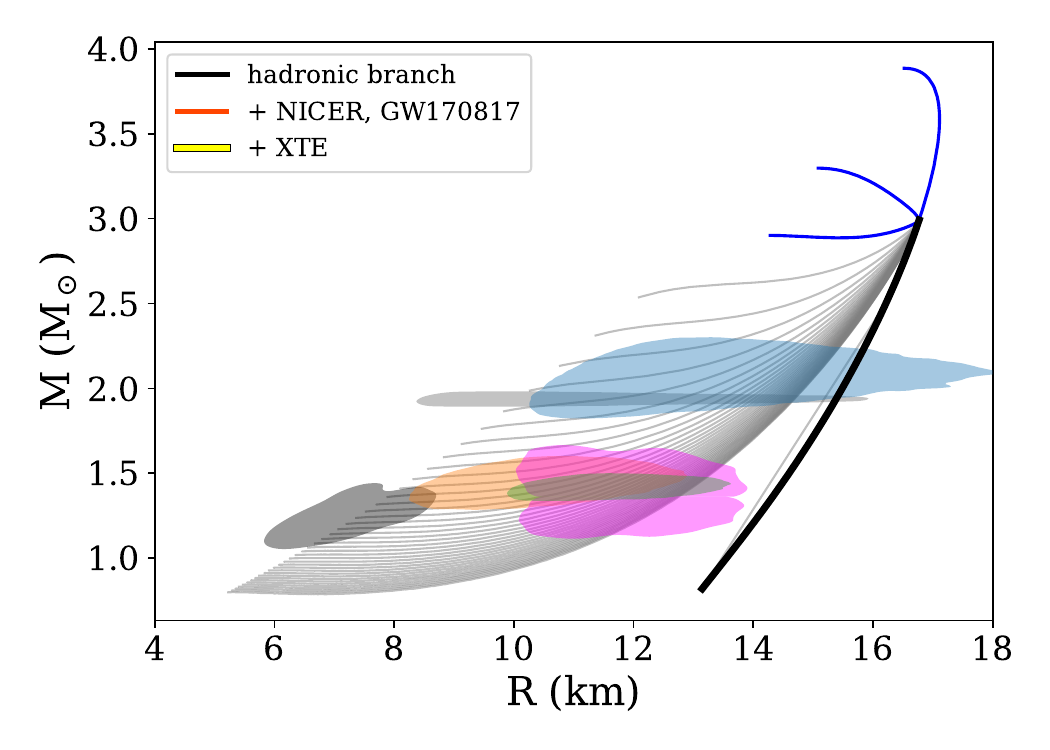}
\caption{EOS (left) and mass--radius relation (right) for the soft branch, $\Gamma_{\rm CET}=1.77$, at $\mu_{\rm tr}=1000$, $1200$ and $1300$~MeV (top to bottom). In the EOS panels the red and green dashed curves are the boundaries of the allowed region in the $(\varepsilon,p)$ plane, obtained for $\Gamma_{\rm CET}=1.77$ with $c_s^2\to0$ and for $\Gamma_{\rm CET}=3.23$ with $c_s^2\to1$ above the matching density, and the dark-blue dashed curves with the shaded region show the pQCD region. Grey EOSs are rejected and coloured EOSs are accepted. In the $M$--$R$ panels, grey curves are EOSs that are pQCD-incompatible or have no stable hybrid branch, and blue curves are accepted EOSs that satisfy the pQCD constraint. Red curves in addition intersect the contours labelled NICER (J0740, J1614, J0437 and J0614) and GW170817, and yellow curves also intersect the contour labelled XTE (J1814-338). The black curve is the hadronic branch common to all jumps. The shaded regions are the observational contours.}
\label{fig:results_soft}
\end{figure*}

Figure~\ref{fig:results_soft} shows the accepted EOSs (left) and the corresponding mass--radius relations (right) for the soft polytrope ($\Gamma_{\rm CET}=1.77$) at $\mu_{\rm tr}=1000$, $1200$ and $1300$~MeV. The number of accepted EOSs, and their classification, are collected in Table~\ref{tab:accepted}, which also contains the intermediate case $\Gamma_{\rm CET}=2.23$ and $\mu_{\rm tr}=1100$ and $1250$~MeV.

\begin{table*}
\caption{Outcome of the scan of $41$ density jumps for each $\Gamma_{\rm CET}$ and $\mu_{\rm tr}$. $n_{\rm tr}$ is the baryon density at the onset of the transition. An EOS has a stable hybrid branch if the mass rises by at least $0.01\,\Msun$ above the transition. ``pQCD-incompatible'' counts the EOSs with a stable hybrid branch that violate the pQCD constraint at the last stable star. EOSs without a stable hybrid branch are not excluded by pQCD and appear in neither column; at $\mu_{\rm tr}=1300$~MeV they are the difference between $41$ and the sum of the two columns. ``Accepted'' counts the EOSs with a stable hybrid branch that satisfy the pQCD constraint, and these are split into three classes, with the colours of Fig.~\ref{fig:results_soft}: blue, pQCD-compatible only; red, pQCD-compatible and consistent with the observations of J0740, J1614, J0437, J0614 and GW170817; yellow, pQCD-compatible and consistent with these observations and with J1814-338. An EOS is consistent with an observation if its $M$--$R$ curve passes through the corresponding contour. $M_{\rm max}$ is the range of maximum masses of the accepted EOSs.}
\label{tab:accepted}
\centering
\small
\begin{tabular}{ccccccccc}
\toprule
$\Gamma_{\rm CET}$ & $\mu_{\rm tr}$ (MeV) & $n_{\rm tr}/n_0$ & accepted & blue & red & yellow & pQCD-incompatible & $M_{\rm max}$ ($\Msun$)\\
\midrule
1.77 & 1000 & 1.15 & 17 & 14 & 3 & 0 & 24 & 1.32--3.89\\
1.77 & 1100 & 1.27 & 19 & 19 & 0 & 0 & 22 & 1.70--3.89\\
1.77 & 1200 & 1.39 & 22 & 20 & 0 & 2 & 19 & 2.48--3.89\\
1.77 & 1250 & 1.44 & 23 & 16 & 0 & 7 & 18 & 2.76--3.89\\
1.77 & 1300 & 1.50 & 3 & 3 & 0 & 0 & 3 & 2.98--3.89\\
\addlinespace
2.23 & 1000 & 1.15 & 17 & 14 & 3 & 0 & 24 & 1.32--3.89\\
2.23 & 1100 & 1.26 & 19 & 18 & 1 & 0 & 22 & 1.69--3.89\\
2.23 & 1200 & 1.38 & 22 & 20 & 0 & 2 & 19 & 2.48--3.89\\
2.23 & 1250 & 1.43 & 23 & 17 & 0 & 6 & 18 & 2.76--3.89\\
2.23 & 1300 & 1.49 & 3 & 3 & 0 & 0 & 3 & 2.98--3.89\\
\addlinespace
3.23 & 1000 & 1.12 & 17 & 14 & 3 & 0 & 24 & 1.33--3.93\\
3.23 & 1100 & 1.23 & 19 & 18 & 1 & 0 & 22 & 1.68--3.93\\
3.23 & 1200 & 1.34 & 22 & 19 & 1 & 2 & 19 & 2.48--3.93\\
3.23 & 1250 & 1.40 & 23 & 20 & 0 & 3 & 18 & 2.76--3.93\\
3.23 & 1300 & 1.45 & 3 & 3 & 0 & 0 & 8 & 2.99--3.93\\
\bottomrule
\end{tabular}
\end{table*}

The accepted EOSs are those with a stable hybrid branch that satisfy the pQCD constraint at the last stable star (Table~\ref{tab:accepted}). The largest accepted jump is the same for the three polytropes, $\Delta n=1.28$, $1.44$, $1.68$ and $1.76~{\rm fm^{-3}}$ for $\mu_{\rm tr}=1000$, $1100$, $1200$ and $1250$~MeV, and larger jumps violate the pQCD constraint. At $\mu_{\rm tr}=1300$~MeV only the three smallest jumps ($\Delta n\le0.16~{\rm fm^{-3}}\simeq1\,n_0$) are accepted. Most of the remaining EOSs there, $35$ for $\Gamma_{\rm CET}=1.77$ and $2.23$ and $30$ for $3.23$, have no stable hybrid branch. This is not a pQCD constraint, since such EOSs are not excluded by pQCD, and their maximum-mass star is the last hadronic star. The few other EOSs at $1300$~MeV ($3$, $3$ and $8$) violate the pQCD constraint. In the EOS panels the transition appears as a horizontal segment at constant pressure, every accepted curve remains causal ($c_s^2\le1$), and all curves end inside the pQCD envelope by construction. Increasing $\Delta n$ lowers the pressure at a given energy density in the intermediate-density range, so the spread among the accepted EOSs is widest there and narrows towards the pQCD region.

The maximum mass depends strongly on $\mu_{\rm tr}$. For $\mu_{\rm tr}\ge1200$~MeV every accepted EOS supports $\MTOV>2\,\Msun$, whereas for $\mu_{\rm tr}=1000$~MeV only $35\%$ (6 of 17) do, since large jumps at low chemical potential soften the EOS too strongly. The upper end of the mass range ($\simeq3.9\,\Msun$) is populated by the stiffest members of the ensemble ($\Delta n\to0$, $c_s^2=1$ beyond the polytrope), which lie close to the causal limit of Sect.~\ref{sec:crust}; we do not impose an upper bound on $\MTOV$ here.

The radii are controlled by the position of the transition. For $\mu_{\rm tr}\ge1200$~MeV a $1.4\,\Msun$ star on the hadronic branch lies below the transition, so its radius is fixed by the polytrope alone, $R_{1.4}=14.40$~km (soft) and $14.91$~km (stiff), independent of $\Delta n$. These values lie above the radii preferred by the NICER and GW170817 constraints, so compatibility must come from the hybrid branch. For $\mu_{\rm tr}=1200$~MeV this happens through twin-like behaviour: $11$ of the $22$ accepted curves, for each polytrope, cross $M=1.4\,\Msun$ at two or more radii, for jumps $\Delta n=0.88$--$1.68~{\rm fm^{-3}}$ ($5.5$--$10.5\,n_0$). Besides the hadronic radius, $14.4$~km (soft) or $14.9$~km (stiff), the crossing on the hybrid branch lies at $8.6$--$13.2$~km (soft) and $8.5$--$13.6$~km (stiff), corresponding to a hadronic and a hybrid star of the same mass. Smaller jumps leave $R_{1.4}$ unchanged. Since a stable hybrid segment is part of the acceptance criterion, its existence alone is not a result; the crossing at fixed mass is.

% \subsubsection{Compatibility with observations}
% \label{sec:sosi_obs}

We classify every accepted curve by the observational regions that it intersects, where a curve intersects a region if any of its $(R,M)$ points lies inside the contour (the $2\sigma$ contours of J0740, J0614, J0437 and both components of GW170817, and the $99\%$ contours of J1614 and J1814-338). Points from failed TOV integrations ($R<1$~km or $M<0.05\,\Msun$) are discarded. The classes are: EOSs that are pQCD-incompatible or have no stable hybrid branch (grey), pQCD-compatible only (blue), pQCD-compatible and intersecting J0740, J1614, J0437, J0614 and both GW170817 contours (red), and additionally J1814-338 (yellow). In Fig.~\ref{fig:results_soft} the observational contours are drawn in front of the curves.

Table~\ref{tab:accepted} shows that the classes are populated in a pattern determined by $\mu_{\rm tr}$ rather than by $\Gamma_{\rm CET}$. At $\mu_{\rm tr}=1000$~MeV three curves per $\Gamma_{\rm CET}$ are red (for $\Gamma_{\rm CET}=1.77$: $\Delta n=0.16$, $0.24$ and $0.32~{\rm fm^{-3}}$), but none reaches J1814-338: the curves that do reach this compact region ($0.56\le\Delta n\le1.28~{\rm fm^{-3}}$ for $\Gamma_{\rm CET}=1.77$) miss J0740, J0437, J0614 and GW170817. Among the values of $\mu_{\rm tr}$ tested, yellow curves, which agree with all six observations, appear only at $\mu_{\rm tr}=1200$ and $1250$~MeV: two for each $\Gamma_{\rm CET}$ at $1200$~MeV, and $7$, $6$ and $3$ for $\Gamma_{\rm CET}=1.77$, $2.23$ and $3.23$ at $1250$~MeV. None appears at $1000$, $1100$ or $1300$~MeV, so the transition has to lie at intermediate chemical potentials, between $1100$ and $1300$~MeV, for the EOS to be consistent with all six observations. Our scan, with a spacing of $50$--$100$~MeV, does not resolve the allowed range further. Of the values tested the overlap is best at $1250$~MeV, and the soft polytrope is more compatible than the stiff one. The $\mu_{\rm tr}=1250$~MeV case is listed in Table~\ref{tab:accepted} but not shown in Fig.~\ref{fig:results_soft}.

The onset densities of the EOSs in this section, $n_{\rm tr}\simeq1.1$--$1.5\,n_0$ (Table~\ref{tab:accepted}), lie close to the window $n_m<\ntr<\ntr^*$ of Sect.~\ref{sec:crust}, for the matching density $n_m=1.1\,n_0$ that is also used here. For $M_{\rm obs}=2.01\,\Msun$ that window ends at $\ntr^*\simeq1.27\,n_0$ and the causal bound is $\dmax\simeq460$--$494~{\rm MeV\,fm^{-3}}$ (Table~\ref{tab:bounds}). Two of our five values of $\mu_{\rm tr}$ give onsets inside the window, $\mu_{\rm tr}=1000$ ($\ntr\simeq1.12$--$1.15\,n_0$) and $1100$~MeV ($1.23$--$1.27\,n_0$), and for them we can compare the pQCD-matched EOSs with the bound. For $\Gamma_{\rm CET}=3.23$, the polytrope of Sect.~\ref{sec:crust}, the onsets are $\ntr=1.119$ and $1.230\,n_0$, where the bound is $460$ and $482~{\rm MeV\,fm^{-3}}$. The jump is sampled in steps of $0.08~{\rm fm^{-3}}$, which by Eq.~(\ref{eq:deps_dn}) corresponds to $\Delta\ve=80$ and $88~{\rm MeV\,fm^{-3}}$ at $\mu_{\rm tr}=1000$ and $1100$~MeV, so that the limit of the pQCD-matched EOSs is only determined to within one step. At $1000$~MeV the EOSs with $\Delta n=0.40~{\rm fm^{-3}}$ ($\Delta\ve=400~{\rm MeV\,fm^{-3}}$) still support $\MTOV\ge2.01\,\Msun$, while those with $0.48~{\rm fm^{-3}}$ ($480~{\rm MeV\,fm^{-3}}$) do not, and at $1100$~MeV the corresponding interval is $440$--$528~{\rm MeV\,fm^{-3}}$. The causal bounds lie inside these intervals, as they must, and the softer polytropes behave in the same way. This is a consistency check of the construction of the EOSs of this section against the extremal EOS of Sect.~\ref{sec:crust}; its resolution is set by the grid in $\Delta n$. For $\mu_{\rm tr}\ge1200$~MeV, $n_{\rm tr}\ge1.34\,n_0$ exceeds $\ntr^*$, so the hadronic branch alone supports $2.01\,\Msun$ and no bound is expected. Accordingly, all accepted EOSs at $\mu_{\rm tr}\ge1200$~MeV have $\MTOV>2\,\Msun$ irrespective of the jump, up to $\Delta n\simeq1.8~{\rm fm^{-3}}$ ($\Delta\ve\simeq2.2\times10^{3}~{\rm MeV\,fm^{-3}}$).

%=============================================================================
\subsection{Stable twin branches and the compact source J1814-338}
\label{sec:twins}
%=============================================================================

\begin{figure}
\centering
\includegraphics[width=\columnwidth]{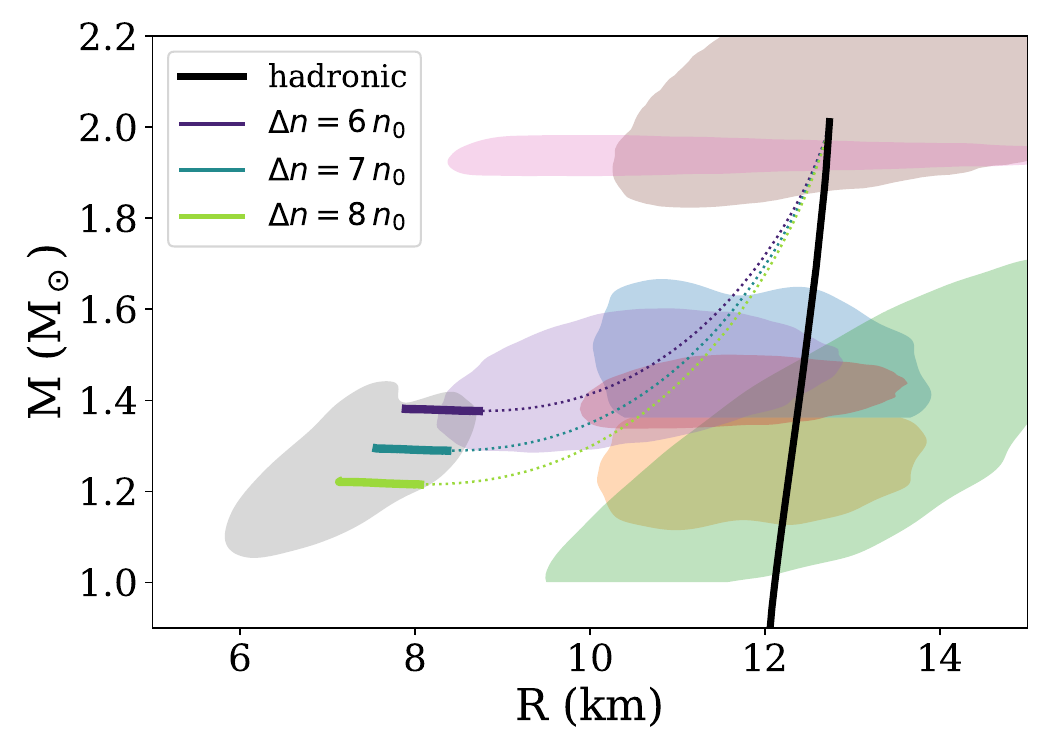}
\caption{Stable mass--radius stretches for the permissible EOS with the lowest break density, with a first-order transition inserted where the hadronic $M$--$R$ curve reaches $2.01\,\Msun$, for density jumps $\Delta n=6$, 7 and $8\,n_0$ (the $c_s^2=1$ continuation is used above the jump). The black curve is the hadronic branch, which is common to all jumps and ends at $M=2.01\,M_\odot$, $R=12.74$~km. Dotted: the hybrid sequence from the transition to the start of the stable stretch. Solid: the stable stretch, from the valley to the maximum of the mass; the remainder of the hybrid sequence is shown in Fig.~\ref{fig:app_197_full}. The shaded regions are the observational contours.}
\label{fig:stable_j1814_197}
\end{figure}

The EOSs of Sect.~\ref{sec:sosi} are accepted or rejected as a whole, and their stellar sequences are fixed by $(\mu_{\rm tr},\Delta n)$. To test whether a stable hybrid branch can reproduce the compact source J1814-338 in addition to the heavier pulsars, we now work with an ensemble of agnostically constructed hadronic EOSs \cite{saha_2026_20502449}. The methodology for constructing such EOSs is described in \ref{app:ensemble}. From this ensemble we first select the EOSs whose $M$--$R$ curves satisfy the observations of PSR J0030, J0437, J0614, J0740 and J1614 simultaneously, and we call these the permissible EOSs. For each permissible EOS we then cut the EOS at the central energy density where the hadronic $M$--$R$ curve reaches $M=2.01\,\Msun$, the mass for which the causal bound of Sect.~\ref{sec:crust} is evaluated. The part of the EOS below the cut passes through the contours of the observations listed above, and the stitched EOS supports at least $2.01\,\Msun$. A first-order transition is stitched on at this point, and we scan the density jump to see which EOSs can satisfy the XTE observation of J1814-338. The transition is added at constant $p$ and $\mu$, with the baryon density jumping by $\Delta n=Z\,n_0$ and the energy density by $\Delta\varepsilon=\mu_{\rm tr}\,\Delta n$, as in Eq.~(\ref{eq:deps_dn}). Above the jump the EOS is continued with $c_s^2=1$,
\begin{equation}
\varepsilon=\varepsilon_{\rm post}+(p-p_{\rm tr}),\qquad \varepsilon_{\rm post}=\varepsilon_{\rm tr}+\Delta\varepsilon ,
\end{equation}
and the TOV equations are solved for central energy densities below and above the jump. We show the permissible EOS with the lowest break density. For this EOS the hadronic $M$--$R$ curve reaches $2.01\,\Msun$ at $\varepsilon_c=463~{\rm MeV\,fm^{-3}}$, corresponding to $n=2.82\,n_0$, $p=115~{\rm MeV\,fm^{-3}}$ and $\mu_{\rm tr}=(\varepsilon+p)/n=1279$~MeV, where the star has $R=12.74$~km.

Stability is decided by the sign of $dM/d\varepsilon_c$. The hadronic branch has $dM/d\varepsilon_c>0$ at every point up to the transition. On the hybrid branch the mass first drops steeply after the jump and then flattens into a shallow valley, followed by a small maximum, before it decreases slowly again. We identify the largest such rise on the unsmoothed sequence, and the valley-to-maximum stretch with $dM/d\varepsilon_c>0$ is the stable hybrid branch. The rises are small, $0.004$--$0.006\,M_\odot$, and we therefore require a net rise of at least $0.001\,M_\odot$, which exceeds the numerical noise: for $\Delta n=6\,n_0$ the rise changes only from $0.0044$ to $0.0042\,M_\odot$ when the RK4 step is reduced from $1$ to $0.1$~m. The branch beyond the maximum is unstable.

Figure~\ref{fig:stable_j1814_197} shows the result for $\Delta n=6$, 7 and $8\,n_0$ ($\Delta\varepsilon=1228$, 1432 and $1637~{\rm MeV\,fm^{-3}}$). In each case the hybrid branch is disconnected from the hadronic one, so that stars of the same mass exist on both branches, with $R\simeq12.7$~km on the hadronic branch and a considerably smaller radius, $R\simeq7$--$9$~km, on the hybrid branch (stable twins). The properties of the stable stretch are listed in Table~\ref{tab:stable_197}. Increasing $\Delta n$ moves the stretch to lower masses and smaller radii, $M\simeq1.38$, $1.29$ and $1.22\,M_\odot$ for $\Delta n=6$, 7 and $8\,n_0$. For all three jumps the stretch is compatible with J1814-338, so that a single EOS describes both the massive stars up to $2.01\,\Msun$ that pass through the contours of J1614 and J0740 and the compact J1814-338. The stable stretch is short: the mass changes by only $0.004$--$0.006\,M_\odot$ over a range of central densities of about $700$--$1100~{\rm MeV\,fm^{-3}}$, so that it is a narrow stable window and not an extended branch. For $\Delta n=5\,n_0$ a stable window also exists, at $M=1.48\,M_\odot$, but it is not compatible with J1814-338, since it lies above the mass range of the observation (Fig.~\ref{fig:app_197_full}). All four windows are compatible with pQCD. At the last stable star, with $\varepsilon_c=2837$, $3326$, $3743$ and $4394~{\rm MeV\,fm^{-3}}$ for $\Delta n=5$, 6, 7 and $8\,n_0$, the pQCD constraint of \ref{app:construction} is satisfied for renormalisation-scale factors $0.54\le X\le2$, $0.59\le X\le2$, $0.68\le X\le2$ and $0.82\le X\le2$, where $X$ is varied between $1/2$ and $2$; the constraint is also satisfied at the start of each window.

\begin{table}
\caption{Stable stretch of the hybrid branch of the permissible EOS with the lowest break density, with the transition at $2.01\,\Msun$, from the valley to the maximum of the mass. $\varepsilon_c$ is the central energy density at the valley and at the maximum. The last column states whether the stretch intersects the J1814-338 contour.}
\label{tab:stable_197}
\centering
\footnotesize\setlength{\tabcolsep}{3pt}
\begin{tabular}{cccccc}
\toprule
$\Delta n$ & $\Delta\varepsilon$ & $\varepsilon_c$ range & $M$ range & $R$ range & J1814 \\
$(n_0)$ & (MeV\,fm$^{-3}$) & (MeV\,fm$^{-3}$) & ($M_\odot$) & (km) & \\
\midrule
5 & 1023 & 2242--2837 & 1.480--1.485 & 9.18--8.35 & no \\
6 & 1228 & 2615--3326 & 1.377--1.381 & 8.73--7.90 & yes \\
7 & 1432 & 2972--3743 & 1.290--1.294 & 8.37--7.56 & yes \\
8 & 1637 & 3328--4394 & 1.215--1.221 & 8.06--7.13 & yes \\
\bottomrule
\end{tabular}
\end{table}

The existence of the stable window depends on the position of the transition and on the hadronic EOS. For the permissible EOS with the lowest break density the window is much thinner than for a transition at a lower mass: cutting the same EOS at the first entry into the J1614 contour, $M=1.906\,M_\odot$ ($\varepsilon_c=440~{\rm MeV\,fm^{-3}}$), gives a net rise of $0.04$--$0.05\,M_\odot$ instead of $0.004$--$0.006\,M_\odot$. For the three permissible EOSs with higher break densities shown in \ref{app:twins} (Fig.~\ref{fig:app_nostable}), whose hadronic curves reach $2.01\,M_\odot$ at $\varepsilon_c=602$, $654$ and $695~{\rm MeV\,fm^{-3}}$, the net rise is at most $0.0002\,M_\odot$ for $\Delta n=5$--$8\,n_0$, which is at the level of the numerical noise, and no stable window forms. We note that the stability test uses the raw mass sequence without smoothing, and the threshold of $0.001\,M_\odot$ is a choice, whose dependence on the integration step was tested only for $\Delta n=6\,n_0$. The jumps tested are $\Delta n=5$--$8\,n_0$, and we have not explored the dependence on the sound speed above the jump.

%=============================================================================
\section{Discussion and Summary}
\label{sec:discussion_summary}
%=============================================================================

We have generalised the causality bound on the compactness and the maximum mass of NSs to EOSs with a FOPT. In units of the onset energy density, the problem depends only on $q=\ptr/\etr$ and $j=\Delta\ve/\etr$. The smooth bound $\mathcal{C}\le0.354$ is recovered exactly for $q\to0$ or $j\to0$, and the transition has no effect for $q>0.670$. Elsewhere the maximum compactness is reduced by typically $\lesssim0.04$ when the heaviest star is a hybrid star, and by up to $0.13$ when it is the last hadronic star.

Matching to a BPS crust and a CET polytrope at $n_m=1.1\,n_0$ and inverting the bound gives upper limits on the jump that hold for any causal EOS above the transition, since they rely only on causality, the low-density EOS and the observed mass (and radius). A $2.01\,\Msun$ star excludes $\Delta\ve\gtrsim460$--$495~{\rm MeV\,fm^{-3}}$ for onsets just above $n_m$, and a radius of at least $11.6$~km lowers this to $\simeq235$--$395~{\rm MeV\,fm^{-3}}$. The limits depend only weakly on the polytrope index: at $\ntr=1.12\,n_0$ we find $\dmax=456$, $457$ and $461~{\rm MeV\,fm^{-3}}$ for $\Gamma_{\rm CET}=1.77$, $2.23$ and $3.23$. They apply only for onsets below $\ntr^*$, which lies $\simeq0.2\,n_0$ above $n_m$. Above $\ntr^*$ a heavy star may be purely hadronic, and mass measurements alone do not constrain the transition. On the stiffest hadronic branch such a star is large, $R\simeq15.9$~km at $2.01\,\Msun$, which exceeds the NICER estimates for PSR J0740+6620~\cite{Riley2021,Miller2021,Salmi2024}. Turning this into a bound requires an upper radius limit and an optimisation over the hadronic segment, which we leave to future work. We have also assumed rapid conversion at the interface. For slow conversion, hybrid configurations with $dM/d\varepsilon_c<0$ can remain stable~\cite{Pereira2018}, which would modify region~II of Fig.~\ref{fig:dbeta} and the twin-star branches, but not the mass bound, which uses only the maximum mass.

We have also built EOSs with a first-order transition at $\mu_{\rm tr}=1000$--$1300$~MeV and a speed-of-sound interpolation matched to pQCD at $2.6$~GeV. In our scan of values for $\mu_{\rm tr}$, the pQCD constraint limits the jump to $\Delta n\le1.28$--$1.76~{\rm fm^{-3}}$ for $\mu_{\rm tr}=1000$--$1250$~MeV. At $1300$~MeV most jumps have no stable hybrid branch, which is a property of the EOS and not a pQCD constraint. EOSs compatible with J0740, J1614, J0437, J0614, GW170817 and J1814-338 appear only at $1200$ and $1250$~MeV, i.e.\ for transitions at intermediate chemical potentials, between $1100$ and $1300$~MeV. For the onsets inside the window of the causal bound, the largest jump compatible with $2.01\,\Msun$ is consistent with the bound within the resolution of the jump grid. This is expected, because the sound speed inside stable stars is $c_s^2=1$ on both sides of the transition in these EOSs. The observational classification uses only the intersection of a curve with each contour, not a likelihood, and the jump is sampled in steps of $0.5\,n_0$, so the counts in Table~\ref{tab:accepted} outline the allowed region and are not probabilities.

Finally, we started from an ensemble of agnostically constructed hadronic EOSs and selected those that satisfy PSR J0030, J0437, J0614, J0740 and J1614 simultaneously. We cut each EOS at the density where its hadronic $M$--$R$ curve reaches $2.01\,\Msun$, the mass used for the causal bound, and stitched on a first-order transition with $c_s^2=1$ above the jump. For the permissible EOS with the lowest break density this gives a narrow stable window of the hybrid branch, with a mass change of only $0.004$--$0.006\,M_\odot$ over central densities of $700$--$1100~{\rm MeV\,fm^{-3}}$, which passes through the J1814-338 contour for jumps of $\Delta n=6$--$8\,n_0$, and the EOS satisfies the pQCD constraint at the end of the window. The window is much thinner than for a transition at $1.906\,M_\odot$, and it is absent for three other permissible EOSs, whose breaks at $2.01\,M_\odot$ lie at $602$--$695~{\rm MeV\,fm^{-3}}$. The jumps are far above $\dmax$, but there is no conflict with the causal bound: the transition starts at $n=2.8\,n_0$, well above $\ntr^*$, so that the observed masses are carried by the hadronic branch and the hybrid branch only has to reproduce J1814-338.

\begin{acknowledgements}
The authors thank IISER Bhopal for providing the infrastructure to carry out the research. RM is grateful to the Science and Engineering Research Board (SERB), Govt.\ of India, for monetary support in the form of Core Research Grant (CRG/2022/000663).
\end{acknowledgements}

\section*{Data Availability Statement} This manuscript has no associated data. The numerical codes used to generate all results and figures are available from the authors on reasonable request.

\section*{Code Availability Statement} The codes will be made available on reasonable request.

%========================================================================================
\bibliographystyle{spphys}
\bibliography{ref}
%=============================================================================

\appendix
\section{Construction of the EOSs of Sect.~\ref{sec:sosi}}
\label{app:construction}

This appendix describes how the EOSs of Sect.~\ref{sec:sosi} are constructed. The chemical potential is $\mu=(\ve+p)/n$.

\paragraph{Low-density and hadronic branch.} Below $n_m=1.1\,n_0$ we use the BPS crust~\cite{Baym1971}, continued by the chiral-EFT polytrope of Sect.~\ref{sec:formalism},
\begin{equation}
p(n)=K\left(\frac{n}{n_0}\right)^{\Gamma_{\rm CET}},\qquad n_0=0.16~{\rm fm^{-3}},
\end{equation}
with $K$ fixed by pressure continuity at the upper edge of the crust. We consider a soft ($\Gamma_{\rm CET}=1.77$), an intermediate ($2.23$) and a stiff ($3.23$) polytrope. Above $n_m$ the hadronic branch is continued causally, at $c_s^2=1$,
\begin{equation}
p(\ve)=p_m+(\ve-\ve_m),\qquad \ve\ge\ve_m ,
\end{equation}
exactly as in Eq.~(\ref{eq:eos}). This branch is carried up to the transition chemical potential $\mu_{\rm tr}$, which we take as a free input.

\paragraph{First-order transition.} At $\mu_{\rm tr}$ the pressure and the chemical potential are continuous while the baryon density jumps by $\Delta n$,
\begin{equation}
p^{q}_{\rm tr}=p^{h}_{\rm tr},\qquad \mu^{q}_{\rm tr}=\mu^{h}_{\rm tr},\qquad n^{q}_{\rm tr}=n^{h}_{\rm tr}+\Delta n .
\end{equation}
Since $\ve=\mu n-p$ on both sides, the energy-density jump is
\begin{equation}
\Delta\ve=\mu_{\rm tr}\,\Delta n ,
\label{eq:deps_dn}
\end{equation}
which connects this description to the parameters $(\ptr,\etr,\Delta\ve)$ of Sect.~\ref{sec:formalism}. Immediately above the jump the EOS is continued causally once more. The trial EOS, consisting of the hadronic branch, the jump and the causal continuation, is then causal at every density and has no further free parameters.

\paragraph{Slicing at the maximum-mass star.} We solve the TOV equations for this trial EOS over a grid of central energy densities, which gives $M(\ve_c)$, and locate the largest central density $\ve_c^{\max}$ for which $dM/d\ve_c>0$ holds on the branch above the transition. The location is made robust against fixed-step integration noise by smoothing and a confirmation criterion (\ref{app:branch-finder}). A hybrid sequence is accepted only if its mass rises by at least $0.01\,\Msun$ above the transition. The trial EOS is cut at the chemical potential of this point, $\mu_{m,{\rm TOV}}\equiv\mu(\ve_c^{\max})$, offset by a small fixed margin, which defines $\mu_{\rm start}$. Thus the maximum-mass configuration supported by the causal continuation is exceeded before the high-density behaviour is allowed to deviate from $c_s^2=1$.

\paragraph{Speed-of-sound interpolation.} From $\mu_{\rm start}$ onwards the causal continuation is replaced by a branch whose $c_s^2$ is a piecewise-linear function of $\mu$ through the three points
\begin{equation}
(\mu_{\rm start},\,c^2_{s,\rm start}\simeq1),\quad(\mu_{\rm mid},\,c^2_{s,\rm mid}),\quad(\mu_{\rm end},\,c^2_{s,\rm pQCD}),
\end{equation}
with $\mu_{\rm mid}=\alpha\,\mu_{\rm start}$, $\mu_{\rm end}=2.6$~GeV and $c^2_{s,\rm pQCD}=0.33$. The thermodynamic identities
\begin{align}
n(\mu)&=n_{\rm start}\exp\!\left[\int_{\mu_{\rm start}}^{\mu}\frac{d\mu'}{\mu'\,c_s^2(\mu')}\right],\\
p(\mu)&=p_{\rm start}+\int_{\mu_{\rm start}}^{\mu}n(\mu')\,d\mu',
\end{align}
with $\ve=\mu n-p$, then give the EOS. The two free parameters, $\alpha\in[1.00001,1.05]$ and $c^2_{s,\rm mid}\in[10^{-4},0.3]$, are determined jointly by the matching below.

\paragraph{Matching to pQCD.}\label{sec:matching} The pQCD pressure at renormalisation-scale factor $\xi$ is~\cite{Gorda2021,Komoltsev2022}
\begin{equation}
p_{\rm QCD}(\xi,\mu)=\frac{\mu^4}{108\pi^2}\left[c_1-\frac{d_1\,\xi^{-\nu_1}}{\mu-d_2\,\xi^{-\nu_2}}\right],
\end{equation}
with perturbative coefficients $(c_1,d_1,d_2,\nu_1,\nu_2)$, and we scan $\xi\in[0.8,4]$. A trial SoSI branch ends at $(\ve_{\rm end},p_{\rm end})$ at $\mu_{\rm end}$. We invert $p_{\rm QCD}(\xi,\mu_{\rm end})=p_{\rm end}$ for $\xi_{\rm cal}$ and compare $\ve_{\rm end}$ with the pQCD value at that scale,
\begin{align}
\ve_{\rm QCD}&=\mu_{\rm end}\,n_{\rm QCD}-p_{\rm QCD}(\xi_{\rm cal},\mu_{\rm end}),\\ n_{\rm QCD}&=\frac{\partial p_{\rm QCD}}{\partial\mu},
\end{align}
through $\eta=|\ve_{\rm end}-\ve_{\rm QCD}|/\ve_{\rm QCD}$. The pair $(\alpha,c^2_{s,\rm mid})$ minimises $\eta$. The match is accepted only if $\ve_{\rm end}$ lies in the achievable pQCD range at $\mu_{\rm end}$. We obtain this range by scanning $\xi$ directly, since $\ve(\mu_{\rm end};\xi)$ is not monotonic in $\xi$. This match builds the SoSI continuation. We allowed $\xi$ down to $0.8$ in the scan, slightly below the range of the constraint check described next, and EOSs that are accepted by the match but violate the check are counted as pQCD-incompatible.

\paragraph{pQCD constraint.} Whether an EOS is compatible with pQCD is decided by the constraint of Ref.~\cite{Komoltsev2022}. At the last stable star, with energy density $\ve_\ast$, pressure $p_\ast$, baryon density $n_\ast$ and chemical potential $\mu_\ast=(\ve_\ast+p_\ast)/n_\ast$, a causal ($c_s^2\le1$) and thermodynamically consistent path to the pQCD pressure $p_{\rm QCD}$ and density $n_{\rm QCD}$ at $\mu_{\rm end}=2.6$~GeV exists only if
\begin{align}
&\tfrac12\left(\frac{\mu_{\rm end}^2}{\mu_\ast}-\mu_\ast\right)n_\ast < p_{\rm QCD}-p_\ast <\tfrac12\left(\mu_{\rm end}-\frac{\mu_\ast^2}{\mu_{\rm end}}\right)n_{\rm QCD},\\
&n_\ast < n_{\rm QCD}\,\frac{\mu_\ast}{\mu_{\rm end}} .
\end{align}
We evaluate these conditions for $5000$ logarithmically spaced values of the renormalisation-scale factor $X$ between $1/2$ and $2$, which corresponds to $\xi=2X\in[1,4]$ in the notation above, and call the EOS pQCD-compatible if at least one value satisfies them. EOSs with a stable hybrid branch that fail for all $X$ are pQCD-incompatible. EOSs without a stable hybrid branch have no last stable star beyond the transition and are not tested.

\paragraph{Parameter scan.} We scan $41$ values of the density jump, $\Delta n=z\times0.08~{\rm fm^{-3}}$ with $z=0,\dots,40$, i.e.\ $\Delta n\in[0,3.2]~{\rm fm^{-3}}$ in steps of $0.5\,n_0$, for $\mu_{\rm tr}=1000$, $1100$, $1200$, $1250$ and $1300$~MeV and each $\Gamma_{\rm CET}$. The TOV equations are integrated with the RK4 scheme. The hadronic part of the EOS does not depend on $\Delta n$, so all jumps at a given $(\Gamma_{\rm CET},\mu_{\rm tr})$ share one hadronic $M$--$R$ branch.

\section{Bounds for the three observed masses}
\label{app:masses}

\begin{figure}[h]
\centering
\includegraphics[width=\columnwidth]{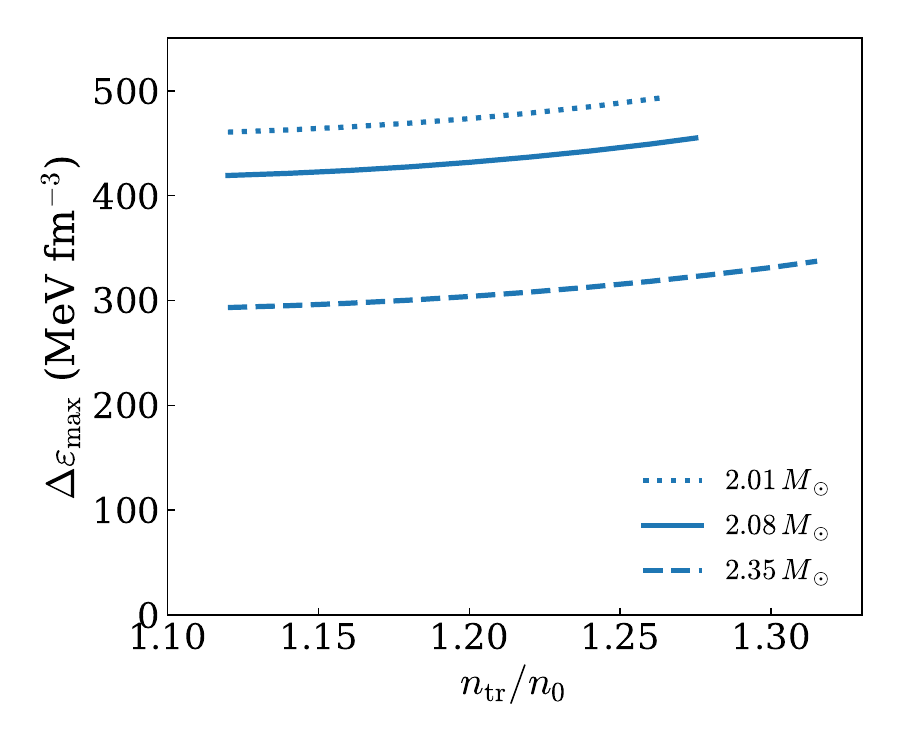}
\caption{Mass-only bound $\dmax(\ntr)$ for $M_{\rm obs}=2.01$ (dotted), $2.08$ (solid) and $2.35\,\Msun$ (dashed), for $n_m=1.1\,n_0$.}
\label{fig:masses}
\end{figure}

Figure~\ref{fig:masses} shows the mass-only bound $\dmax(\ntr)$ of Sect.~\ref{sec:crust} for the three observed masses considered there. A larger observed mass gives a stronger limit and a larger threshold density $\ntr^*$ (Table~\ref{tab:bounds}).

\section{Stable-branch finder}
\label{app:branch-finder}

The $M(\ve_c)$ sequences obtained with a fixed-step integrator contain integration noise, which would produce spurious turning points if $dM/d\ve_c$ were examined point by point. We therefore analyse only the part of the sequence above the transition, after smoothing it with a Savitzky--Golay filter (window of $51$ points, cubic polynomial). Starting right after the jump, we scan towards higher density for the first point where the smoothed mass increases, which is the first local minimum (the valley), where the second (hybrid) branch begins. From there we follow the running maximum of the smoothed mass until it decreases at $5$ consecutive points, which confirms a turn-over, or the grid ends. The stretch from the valley to this maximum is the stable hybrid branch, and its central density at the maximum is $\ve_c^{\max}$ of \ref{app:construction}. The rise of the mass between the valley and the maximum has to reach a threshold, which guards against noise-scale rises: $0.01\,\Msun$ for the EOSs of Sect.~\ref{sec:sosi}, and $0.001\,\Msun$ for the twin-branch EOSs of Sect.~\ref{sec:twins}, whose stable windows are much shorter ($0.004$--$0.006\,\Msun$). The lower threshold is above the numerical noise: for $\Delta n=6\,n_0$ the rise changes only from $0.0044$ to $0.0042\,\Msun$ when the RK4 step is reduced from $1$ to $0.1$~m. If no local minimum is found, or the rise is below the threshold, the sequence is declared to have no stable hybrid branch. Comparing consecutive points directly would miss branches whose rise is spread over many grid points, and we found a real rise of $0.015\,\Msun$ over about $90$ points that point-local tests do not detect. For the twin-branch EOSs, the unsmoothed valley-to-maximum stretch gives the same windows, with masses that agree to $0.0002\,\Msun$ and radii to $0.05$~km.

\begin{figure}[h]
\centering
\includegraphics[width=\columnwidth]{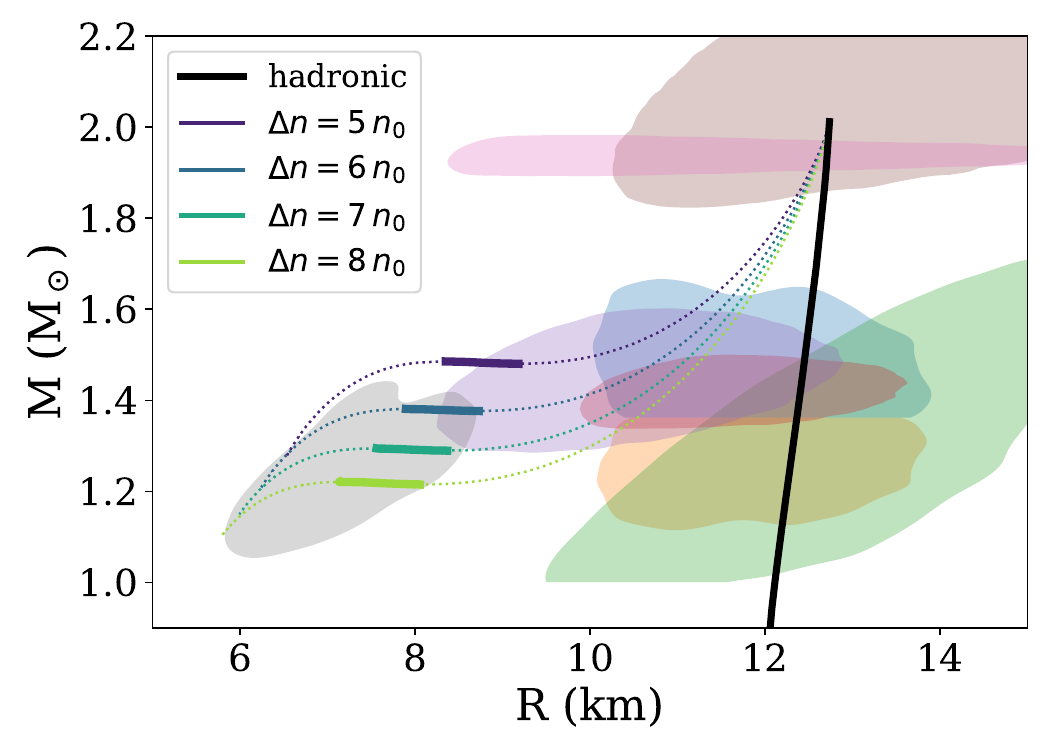}
\caption{The permissible EOS with the lowest break density, for $\Delta n=5$, 6, 7 and $8\,n_0$, with the transition at $2.01\,\Msun$. Solid black: hadronic branch up to the transition. Dotted: the full computed hybrid sequence, including the unstable parts. Solid coloured: the stable stretch from the valley to the maximum of the mass, with a net rise of at least $0.001\,M_\odot$. For $\Delta n=5\,n_0$ the stable stretch lies above the mass range of the J1814-338 contour. The shaded regions are the observational contours as in Fig.~\ref{fig:stable_j1814_197}.}
\label{fig:app_197_full}
\end{figure}

\section{Ensemble EOS construction}
\label{app:ensemble}

The ensemble EOSs of Sect.~\ref{sec:twins} are constructed with the speed-of-sound interpolation method of Ref.~\cite{Annala2020}. The BPS crust~\cite{Baym1971} is followed by a polytrope $p(n)=K(n/n_0)^{\Gamma_{\rm CET}}$ with $\Gamma_{\rm CET}\in[1.77,3.23]$ up to $n_m=1.1\,n_0$. Above $n_m$ the speed of sound is interpolated up to the pQCD anchor at $\mu_{\rm pQCD}=2600$~MeV. The free parameters are $\Gamma_{\rm CET}$ along with $(c_{s,i}^2,\mu_i)$, $i=1,\dots,5$, which gives $11$ parameters in total. Further details are given in Ref. \cite{Saha:2025don}. The EOS of MR shown in Figs. \ref{fig:stable_j1814_197} and \ref{fig:app_197_full} is generated by $\Gamma_{\rm CET}=2.204$, $c_{s,i}^2=\{0.029,\,0.777,\,0.134,\,0.682,\,0.888\}$ and $\mu_i=\{967.4,\,1150.1,\,1650.1,\,1895.4,\,2230.6\}$ MeV.

\section{Full sequences for the twin-branch cases}
\label{app:twins}

\begin{figure*}
\centering
\includegraphics[width=\linewidth]{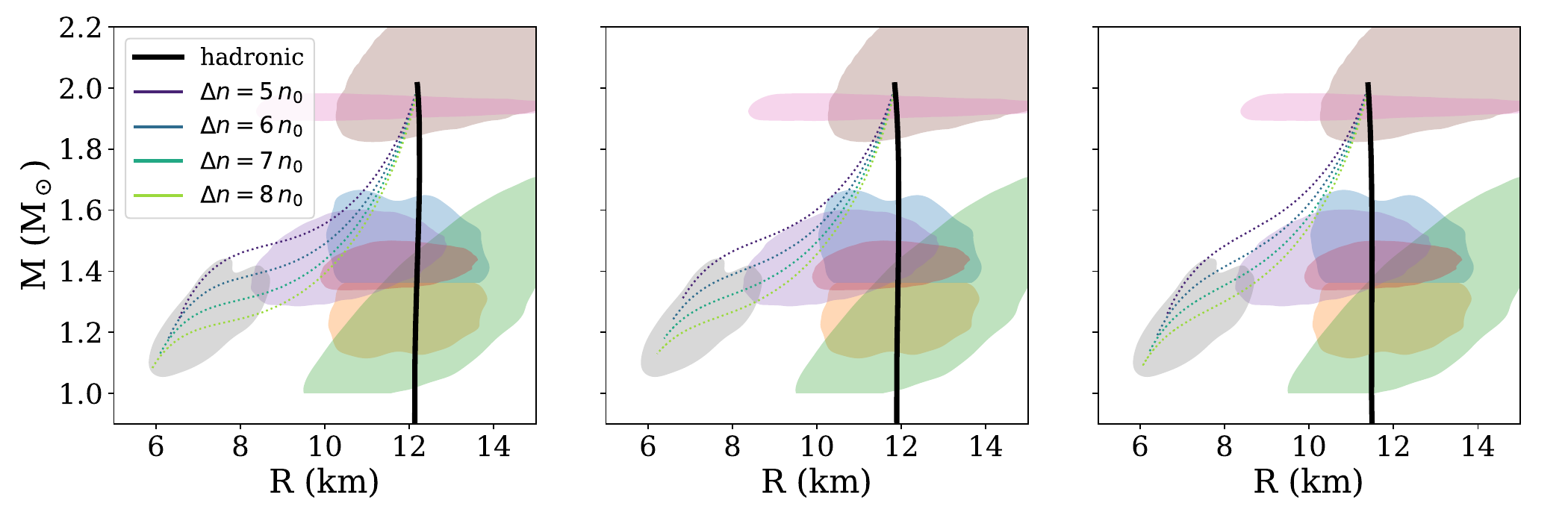}
\caption{Permissible EOSs with higher break densities, whose hadronic $M$--$R$ curve reaches $2.01\,M_\odot$ at $\varepsilon_c=602~{\rm MeV\,fm^{-3}}$ (left), $654$ (middle) and $695$ (right), for $\Delta n=5$, 6, 7 and $8\,n_0$ (colours as in Fig.~\ref{fig:app_197_full}). The hybrid sequences (dotted) develop no stable stretch, since the net rise of the mass is at most $0.0002\,M_\odot$, and only the hadronic branch (solid black) is stable.}
\label{fig:app_nostable}
\end{figure*}

\end{document}